\documentclass[preprint,11pt,showkeys,showpacs,nofootinbib,floatfix]{revtex4}
\usepackage{amssymb}
\usepackage{amsfonts}
\usepackage{mathrsfs}
\usepackage{amsmath}
\usepackage{graphicx}
\usepackage{subfigure}
\usepackage[%dvipdfm,
            pdfstartview=FitH,
            bookmarksnumbered=true,
            bookmarksopen=true,
            colorlinks,
            pdfborder=001,
            linkcolor=blue,
            anchorcolor=green,
            citecolor=red
            ]{hyperref}
\usepackage[toc,page,title,titletoc,header]{appendix}
\usepackage{pdfpages}
\usepackage{booktabs}
\usepackage{multirow}
\begin{document}
\title{Lifshitz scaling effects on holographic paramagnetism/ferromagneism phase transition}
\author{Cheng-Yuan Zhang$^1$}
\author{Ya-Bo Wu$^1$}
\thanks{E-mail address:ybwu61@163.com}
\author{Yong-Yi Jin$^2$}
\author{Yun-Tian Chai$^1$}
\author{Mu-Hong Hu$^1$}
\author{Zhuo Zhang$^1$}
\affiliation{$^1$ Department of physics, Liaoning Normal University, Dalian, 116029, China\\
$^2$ China criminal police University, Shenyang, 110035, China}

%\date{}
%%%%%
\begin{abstract}
In the probe limit, we investigate holographic paramagnetism~-~ferromagnetism phase transition in the four-dimensional (4D) and five-dimensional(5D) Lifshitz black holes by means of numerical and semi-analytical methods, which is realized by introducing a massive 2~-~form field coupled to the Maxwell field. We find that the Lifshitz dynamical exponent $z$ contributes evidently to magnetic moment and hysteresis loop of single magnetic domain quantitatively not qualitatively. Concretely, in the case without external magnetic field, the spontaneous magnetization and ferromagnetic phase transition happen when the temperature gets low enough, and the critical exponent for the magnetic moment is always $1/2$, which is in agreement with the result from mean field theory. And the increasing $z$ enhances the phase transition and increases the DC resistivity which behaves as the colossal magnetic resistance effect in some materials. Furthermore, in the presence of the external magnetic field, the magnetic susceptibility satisfies the Cure-Weiss law with a general $z$. But the increase of $z$ will result in shortening the period of the external magnetic field.
\end{abstract}

\pacs{11.25.Tq, 04.70.Bw, 74.20.-z, 75.20.-g}

\keywords{AdS/CFT correspondence, Holographic ferromagnetism, Lifshitz gravity}
\maketitle
%\tableofcontents

\section{Introduction}
The AdS/CFT correspondence~\cite{Maldacena:1997re,Gubser:1998bc,Witten:1998qj,Witten:1998zw} provides a window into the dynamics of strongly coupled systems by identifying the underlying field theory with a weakly coupled gravity dual. Due to the existence of scaling symmetry near critical point, over the past years the methods and scope of the gauge/gravity have shifted from traditionally QCD-motivated problems to problems in the area of condensed matter systems(see reviews~\cite{Hartnoll:2009sz,Herzog:2009xv,McGreevy:2009xe,Sachdev:2010ch} and references therein) involving the strong interaction. And the duality also gives us a way to understand gravity and condensed matter physics from other side.

One of interesting application of the duality is to study high temperature superconductors that several models of holographic s-wave~\cite{Hartnoll:2008vx,Hartnoll:2008kx} and p-wave superconductors~\cite{Gubser:2008zu,Gubser:2008wv} have been constructed among the various paradigms in condensed matter physics. For example, the holographic s-wave superconductor model was first realized via an Einstein-Maxwell theory coupled to a complex scalar field in a Schwarzschild-AdS black hole background~\cite{Hartnoll:2008vx,Horowitz:2008bn,Horowitz:2010gk,Horowitz:2009ij}. The condensation of the scalar breaks the U(1) symmetry of the system, mimicking the conductor/superconductor phase transition. Sequentially, one analytically studied the superconductor phase transition near the critical point~\cite{Siopsis:2010uq}. Moreover, by an SU(2) gauge field in the bulk, a holographic p-wave superconductor model was constructed~\cite{Gubser:2008wv}, in which the condensed vector field breaks the U(1) symmetry (one of subgroup of SU(2)) as well as the spatial rotational symmetry spontaneously.

Recently, some efforts have been made to generalize the correspondence to systems with less symmetries (see Refs.~\cite{Siopsis:2010uq,Nakamura:2009tf,Donos:2011bh,Horowitz:2012ky,Ling:2013aya,Rozali:2013ama,Cai:2013sua}, for
example) and to the far-from thermal equilibrium problems (see Refs.~\cite{Murata:2010dx,Bhaseen:2012gg,Adams:2012pj,Garcia-Garcia:2013rha,Chesler:2013lia}, for example).

The other application of duality is to study ferromagnetism where the electron spins align to produce a magnetization, which breaks the time reversal symmetry spontaneously and happens in the ferromagnets at the Curie temperature $T_c$ (sometimes it is even higher than the indoor temperature). As we know, magnetism plays a central role in quantum phase transitions and is ubiquitous in many strongly correlated electronic systems, for example, heavy fermion metals. Yet in holographic contexts, due to various technical challenges, models of magnetism are scarce and not extensively explored (see e.g.~\cite{Iqbal:2010eh}).

Ref.~\cite{Cai:2014oca}, a new example of the application of the AdS/CFT correspondence was proposed on the first to understand these challenging systems by realizing the holographic description of the paramagnetism/ferromagnetism phase transition in a dyonic Reissner-Nordstr\"{o}m-AdS black brane. In that model, the magnetic moment is realized by condensation of a real antisymmetric tensor field which couples to the background gauge field strength in the bulk. In the case without external magnetic field, the time reversal symmetry is spontaneously broken and the spontaneous magnetization happens in low temperatures. The critical exponents are in agreement with the ones from mean field theory. In the case of nonzero magnetic field, the model realizes the hysteresis loop of single magnetic domain and the magnetic susceptibility satisfies the Curie-Weiss law. Obviously, this model in Ref.~\cite{Cai:2014oca} give a good starting to explore more complicated magnetic phenomena and quantum phase transitions.

Although a real anti-symmetric tensor field was introduced to realize a holographic magnetic ordered phase in the above model, a more careful analysis shows there is a vector ghost in the model. Hence in Ref.~\cite{Cai:2015bsa} a modified Lagrangian density was put forward, it comes from the dimensional compactification of p-form field in String/M-theory for the anti-symmetric tensor, which is ghost free and causality is well-defined, and keeps all the significant results in the original model qualitatively.

On the basis of Ref.~\cite{Cai:2014oca}, the model was further extended to realize a holographic model of paramagnetism/antiferromagnetism phase transition by introducing two real antisymmetric tensor fields coupling to the background gauge field strength and interacting with each other~\cite{Cai:2014jta}. And then one studied the coexistence and competition of ferromagnetism and p-wave superconductivity by combining a holographic p-wave superconductor model with a holographic ferromagnetism model~\cite{Cai:2014dza}. It was found that the results in Ref.~\cite{Cai:2014dza} depend on the phase appearing firstly (superconductivity or ferromagnetism) besides the self-interaction of magnetic moment of the complex vector field.

On the other hand, for insulator/metal phase transition a gravity duality model was constructed by introducing a massive 2-form field and a dilaton field coupled with U(1) gauge field in asymptotic AdS black brane background. This model shows the colossal magnetoresistance (CMR) effect found in some manganese oxides materials~\cite{Cai:2015wfa}. Further studies based on this model can be discovered, for example, the effect of back reaction on the holographic paramagnetism/ferromagnetism. One found that the phase transition is always second order, which is different from holographic superconductor exhibiting rich phase structures, especially "the retrograde condensate". At present the holographic duality has been applied to two-dimensional magnetic systems~\cite{Iqbal:2010eh,Cai:2014oca,Cai:2015jta}, where the behaviors near the critical temperature were discussed. However, the setup in Ref.~\cite{Yokoi:2015qba} deals with three-dimensional magnetic systems and describes their behaviors in low temperatures where the technology of spintronics is actively developed, besides near the critical temperature. This holographic model in principle can provide a means to analyze phenomena involving magnetization and spin transport, and thus it can introduce new perspectives in the field of spintronics. However, all these holographic ferromagnetic models were constructed only in the relativistic spacetimes. Thus, we wonder whether the above results still hold in nonrelativistic spacetimes, for example, the Lifshitz spacetime, which is our motivation in this paper.

This paper is organized as follows. In section 2, we build a holographic paramagnetism~-~ferromagnetism phase transition model in the Lifshitz black hole with $AdS_{2}$ geometry, which is realized by introducing a massive 2~-~form field coupled to the Maxwell field strength in the bulk. In section 3 by the semi-analytic method we study the magnetic moment and static magnetic susceptibility. The summary and some discussions are included in section 4.

\section{Holographic model}

\subsection{Background}
Recently, the phase transitions in many condensed matter systems are found to be governed by the so-called Lifshitz fixed points which exhibit the anisotropic scaling of spacetime $t \rightarrow b^{z} t$, $\vec{x} \rightarrow b \vec{x}$ ($z\neq 1$), where $z$ is the
Lifshitz dynamical exponent representing the anisotropy of the spacetime. The gravity description dual to this scaling in the D=$d+2$ dimensional spacetime was proposed in Ref.~\cite{Kachru:2008yh}
\begin{equation}\label{metric1}
ds^2=L^2(-r^{2 z} dt^2+r^2 d\vec{x}^2+\frac{dr^2}{r^2}),
\end{equation}
where $d\vec{x}^2=dx_{1}^{2}+\ldots+dx_{d}^{2}$, and $r\in(0,\infty)$. This geometry reduces to the AdS spacetime when $z=1$, while it is a gravity dual with the Lifshitz scaling as $z>1$. The Lifshitz spacetime~\eqref{metric1} can be realized by a massless scalar field coupled to an Abelian gauge field in the following action~\cite{Taylor:2008tg}
\begin{equation}\label{action}
S=\frac{1}{16 \pi G_{d+2}}\int d^{d+2}x \sqrt{-g}(R-2 \Lambda-\frac{1}{2}\partial_{\mu}\varphi\partial^{\mu}\varphi-\frac{1}{4}e^{b\varphi} F_{\mu\nu}F^{\mu\nu}),
\end{equation}
where $\Lambda$ is the cosmological constant, $\varphi$ is a massless scalar and $F_{\mu\nu}$ is an abelian gauge field strength. The generalization of~\eqref{metric1}
to the case with finite temperature is~\cite{Pang:2009ad}
\begin{equation}\label{metric2}
ds^2=L^2(-r^{2 z}f(r) dt^2+\frac{dr^2}{r^2 f(r)}+r^2 \sum_{i=1}^{d} dx_{i}^{2}),
\end{equation}
where
\begin{eqnarray}\label{geometry}
 f(r)&=&1-\frac{r_h^{z+d}}{r^{z+d}},\ \  \Lambda=-\frac{(z+d-1)(z+d)}{2L^2},\label{c34}\\
 \mathcal{F}_{rt}&=&q_{0}r^{z+d-1},\ \ q_{0}^2=2L^2(z-1)(z+d),\ \ e^{b\varphi}=r^{-2d},\ \ b^2=\frac{2d}{z-1}.
\end{eqnarray}
Evidently, choosing the dynamical exponent $z$ to be one reduces the Lifshitz black hole to the Schwarzschild AdS black hole in $(d+2)$~-~dimensions. The Hawking temperature of the black hole is
\begin{equation}\label{HawkingT}
T=\frac{z+d}{4 \pi} r_{h}^{z}.
\end{equation}
where $r_{h}$ denotes the black hole horizon. As we know, some works have been carried out for the influence of the dynamical critical exponent on the properties of holographic superconductors(for details see Ref.~\cite{Brynjolfsson:2009ct,Sin:2009wi,Bu:2012zzb,Fan:2013tga,Abdalla:2013zra,Cai:2009hn,Hartnoll:2012pp}). For example, one studied the scalar condensation in a (3+1)-dimensional Lifshitz black hole background with $z=3/2$ and $z=2$ in Refs.~\cite{Brynjolfsson:2009ct} and~\cite{Sin:2009wi}, respectively. The s-wave and p-wave superconductor models were built in the (3+1)-dimensional Lifshitz black hole with $z=2$~\cite{Bu:2012zzb}. Recently, Abdalla et al in Ref.~\cite{Abdalla:2013zra} investigated the s-wave superconductor phase transition in a three-dimensional Lifshitz black hole in new massive gravity with $z=3$ and found a series of peaks in the conductivity for certain values of the frequency. Based on the previous investigation on the Lifshitz black hole solution, the effects of the Lifshitz dynamical exponent z on the holographic superconductor models were discussed in some detail via numerical and analytical methods, including s-wave and p-wave models with~\cite{Wu:2014dta} or without magnetic field~\cite{Lu:2013tza}~\cite{Wu:2014dta}. Therefore it is interesting to construct holographic ferromagnetic phase transition by using Lifshitz black hole solutions and to study the influences of the dynamical exponent $z$ on the properties of holographic ferromagnetic phase transition and colossal magnetoresistance effect.

\subsection{Model and EoMs}
\label{model}
Following Ref.~\cite{Cai:2015bsa}, below we consider the Lagrangian density consisting of a U(1) field $A_{\mu}$ and a massive 2-form field $M_{\mu\nu}$
in $(d+2)$-dimensional spacetime
\begin{equation}\label{1}
\mathcal{L}_{m}=-F^{\mu\nu}F_{\mu\nu}-\lambda^2(\frac{1}{12}(dM)^2+\frac{m^2}{4}M_{\mu\nu}M^{\mu\nu}+\frac{1}{2}M^{\mu\nu}F_{\mu\nu}+\frac{J}{8}V(M)),
\end{equation}
where $d M$ is the exterior differential of 2-form field $M_{\mu\nu}$, $m^2$ is the squared mass of 2-form field $M_{\mu\nu}$ being greater than zero (see Ref.~\cite{Cai:2015bsa} for detail), $\lambda$ and $J$ are two real model parameters with $J<0$ for producing the spontaneous magnetization, $\lambda^2$ characterizes the back reaction of the 2-form field $M_{\mu\nu}$ to the background geometry and to the Maxwell field strength, and $V(M)$ is a nonlinear potential of the 2-form field describing the self-interaction of the polarization tensor. For simplicity, we take the form of $V(M)$ as follows,
\begin{eqnarray}\label{2}
V(M)&=&(^{*}M_{\mu\nu}M^{\mu\nu})^2=[^{*}(M\wedge M)]^2,
\end{eqnarray}
where $^{*}$ is the Hodge-star operator. As shown in Ref.~\cite{Cai:2015bsa}, this potential shows a global minimum at some nonzero value of $\rho$.

By varying  action \eqref{1}, we can get the equations of motion for the matter fields as
\begin{eqnarray}\label{eqa1}
\nabla^{\tau}(d M)_{\tau\mu\nu}-m^2 M_{\mu\nu}-J(^{*}M_{\tau\sigma}M^{\tau\sigma})(^{*}M_{\mu\nu})&=&F_{\mu\nu},\nonumber\\
\nabla^{\mu}(F_{\mu\nu}+\frac{\lambda^2}{4}M_{\mu\nu})&=&0.
\end{eqnarray}
In what follows, we start to study systematically the effects of the Lifshitz dynamical exponent
$z$ on the holographic ferromagnetic phase transition based on the Lifshitz spacetime~\eqref{metric2} in the probe limit(i.e., neglecting the back reactions of the massive 2-form field to the background Lifshitz geometry and Maxwell field, also including the Maxwell field to the background geometry).
In this probe approximation, the interaction between the electromagnetic response and external field is taken into account so that we can study how spontaneous magnetization influences the electric transport in the following, but they both have little influence on the structures of materials. We take the following self~-~consistent ansatz with matter fields,
\begin{eqnarray}\label{ansatz}
M_{\mu\nu}&=&-p(r)dt\wedge dr+\rho(r)dx\wedge dy,\nonumber\\
A_{\mu}&=&\phi(r)dt+Bx dy,
\end{eqnarray}
where $B$ is a constant magnetic field viewed as the external magnetic field in the boundary field theory.
Thus nontrivial equations of motion in D=$d+2$-dimensional Lifshitz spacetime read,
\begin{eqnarray}\label{eqrhophip}
\rho''+(\frac{f'}{f}+\frac{d+z-3}{r})\rho'-\frac{1}{r^2 f}[m^2+\frac{4 J p^2}{r^{2z-2}}]\rho+\frac{B}{r^2 f}&=&0,\nonumber\\
(m^2-\frac{4 J \rho^2}{r^4})p-\phi'&=&0,\\
\phi''+\frac{d-z+1}{r}\phi'-\lambda^2(\frac{p'}{4}+\frac{(d+1) p}{4r}-\frac{p z}{4r})&=&0,\nonumber
\end{eqnarray}
where a prime stands for the derivative with respect to $r$. In order to solve above equations, we have to specify boundary conditions for the fields. At the horizon $r\rightarrow r_{h}$, we impose $\phi(r_{h})=0$ to satisfy the finite form $A_{\mu}$, while $\rho(r_{h})$ needs to be regular. Near the boundary $r\rightarrow\infty$, the linearized equations give the following asymptotic solution for matter fields,
\begin{eqnarray}\label{asolution}
\rho&=&\rho_{+}r^{\Delta_{+}}+\rho_{-}r^{\Delta_{-}}+\cdots+\frac{B}{m^2},\nonumber\\
\phi&=&\mu-\frac{\sigma}{r^{(d-z)}}+\cdots,~p=\frac{\sigma (d-z)}{m^2 r^{d-z+1}}+\cdots,~~(z\neq d)\\
\phi&=&\mu+\sigma \ln r+\cdots,~~~p=\frac{\sigma}{r m^2}+\cdots.~~(z=d)\nonumber
\end{eqnarray}
where $\Delta_{\pm}=\frac{4-d-z\pm \sqrt{4 (m^2+4)+(d+z) (d+z-8)}}{2}$, $\rho_{\pm}$, $\mu$, and $\sigma$ are all constants. According to gauge/gravity duality and the explanation for the source in Ref.~\cite{Cai:2015bsa}, we treat $\rho_{+}$ as the source of the dual operator, namely, $\rho_{+}=0$, and $\mu$ and $\sigma$ are chemical potential and charge density of dual field theory, respectively.

The Breitenlohner-Freedmeltaan (BF) bound requires $m^2\geq\frac{-(d+z) (d+z-8)}{4}-4$, and the mass squared $m^2$ of massive 2-form field has the lower bound. In this case, there is a logarithmic
term in the asymptotical expansion~\eqref{asolution}. We treat the coefficient of this logarithmic term as the
source which is set to be zero to avoid the instability induced by this term according to~\cite{Horowitz:2008bn}.
Within the BF bound condition, there does not exist the $AdS_2$ geometry, and the near horizon geometry of an extremal black brane with vanishing temperature. To find the restriction to the parameters, let us consider Eqs.~\eqref{eqrhophip} in the high temperature region where $\rho$ vanishes and we can read off the effective mass square of $\rho$ at the horizon as,
\begin{equation}\label{effmass}
m_{eff}^{2}=m^2+4 J p(r_{h})^2=m^2+\frac{4 J \mu^{2} (d-z)^2}{m^4 r_{h}^{2 d}}=m^2+\frac{4 J \mu^2 (d-z)^2}{m^4} (\frac{z+d}{4 \pi T})^{\frac{2 d}{z}}.
\end{equation}
Because of $J<0$, the temperature term contributes a negative term into the effective mass square, which is divergent when $T\rightarrow0$. It follows that whether we choose the grand canonical ensemble or the canonical ensemble, the instability always appears provided that the temperature is low enough.

\subsection{Spontaneous magnetization and susceptibility}
In this paper, we consider the canonical ensemble where the charge density $\sigma$ will be fixed when we discuss the Lifshitz black hole background. Concretely, we firstly consider the cases of $z=1$ and $2$
in the 4D spacetime as examples by the numerical and analytic methods, and then extend to the cases of $z=1$, $2$ and $3$ in the 5D spacetime. Now we think about the spontaneous magnetization in this probe approximation in the 4D spasetime. To begin with, we compute the critical temperature $T_{c}$ when $B=0$. Similar to the case in the Ref.~\cite{Cai:2015bsa}, the polarization field $\rho$ is a small quantity near the critical temperature, we can neglect the nonlinear terms of $\rho$ in the equations of $\phi$ and $p$. Then we can get
\begin{eqnarray}\label{phip}
\phi(r)&=&\mu (1-\frac{1}{r^{2-z}}),~~p(r)=\frac{\mu(2-z)}{m^2 r^{3-z}},~~(z\neq d),\nonumber\\
\phi(r)&=&\mu \ln{r},~~p(r)=\frac{\mu}{m^2 r}. ~~(z=d)
\end{eqnarray}
In the following calculations, we will consider these two cases of $z\neq d$ and $z=d$. Then
the equations of $\rho$ are as follows,
\begin{eqnarray}\label{eqrhoinfinite}
\rho''+(\frac{f'}{f}+\frac{z-1}{r})\rho'-\frac{1}{r^2 f}[m^2+\frac{4 J \mu^2 (2-z)^2}{m^4 r^{4}}]\rho&=&0,~~(z\neq d),\nonumber\\
\rho''+(\frac{f'}{f}+\frac{1}{r})\rho'-\frac{1}{r^2 f}[m^2+\frac{4 J \mu^2}{m^4 r^{4}}]\rho&=&0.~~(z=d)
\end{eqnarray}
To find the critical temperature, at the horizon, the initial conditions are,
\begin{eqnarray}\label{rhohorizon}
\rho'&=&\frac{m^6+4 J \mu^2 z^2-16 J \mu^2 (1-z)}{(z+2) m^4},~~~~~\rho(r_{h})=1.~~(z\neq d)\nonumber\\
\rho'&=&\frac{m^6+4 J \mu^2}{4 m^4},~~~~~\rho(r_{h})=1.~~~(z=d)
\end{eqnarray}
When we perform the numerical computation, we can first fix the horizon radius $r_{h}=1$, thus the temperature is also fixed. Without loss the generality, we take $\rho(r_{h})=1$ and treat the chemical potential $\mu$ as the shooting parameter to match the boundary condition $\rho_{+}=0$. And then we can use the scaling transformations to transform our results into the case in canonical ensemble where the charge density is fixed. As a typical example, we also choose parameters as $m^2=-J=1/8$ and $\lambda=1/2$. Thus the critical temperature $T_{c}$ will be found which are listed in Tab.~\ref{tab:tc31}.
\begin{figure}
\centering
\includegraphics[width=0.4\textwidth]{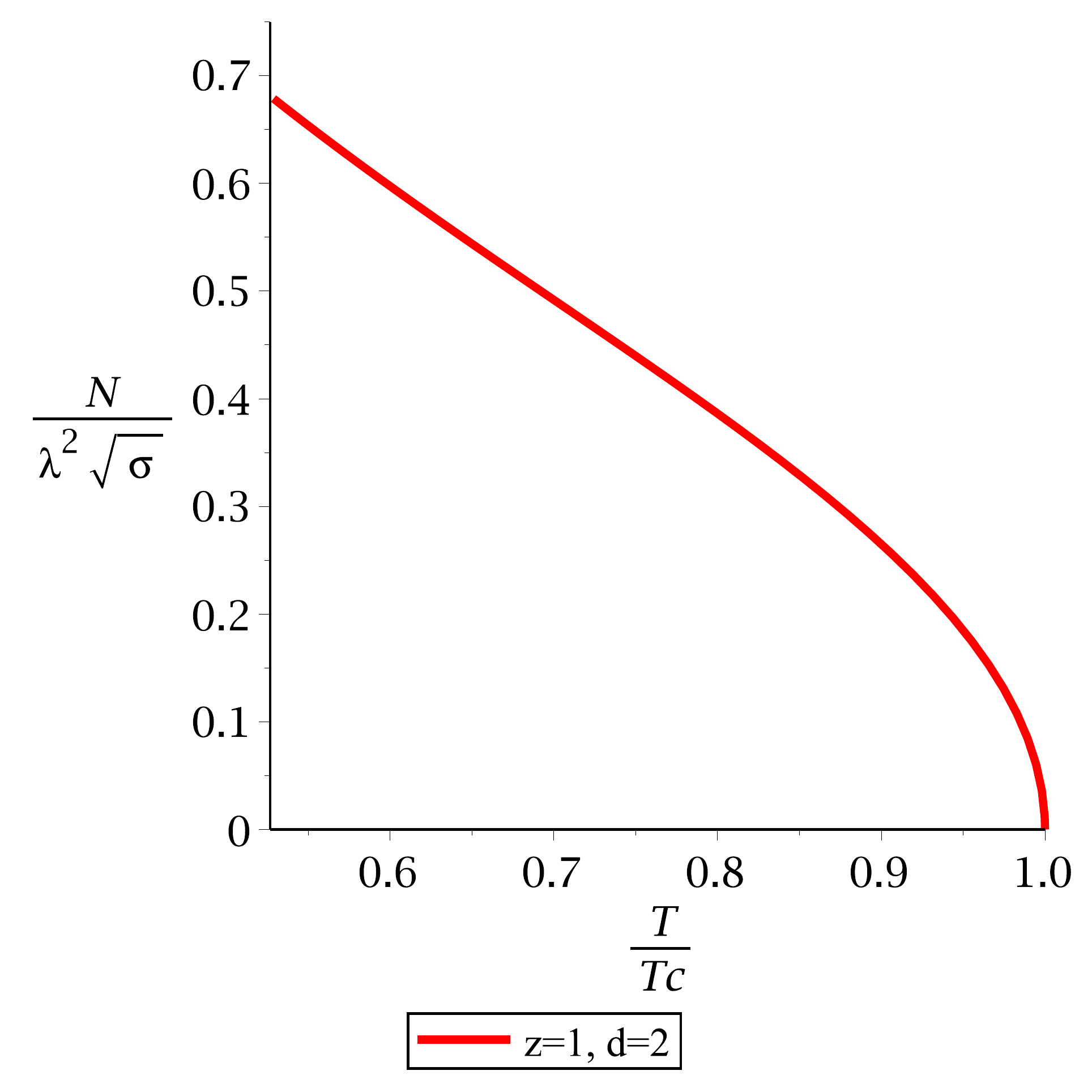}
\includegraphics[width=0.4\textwidth]{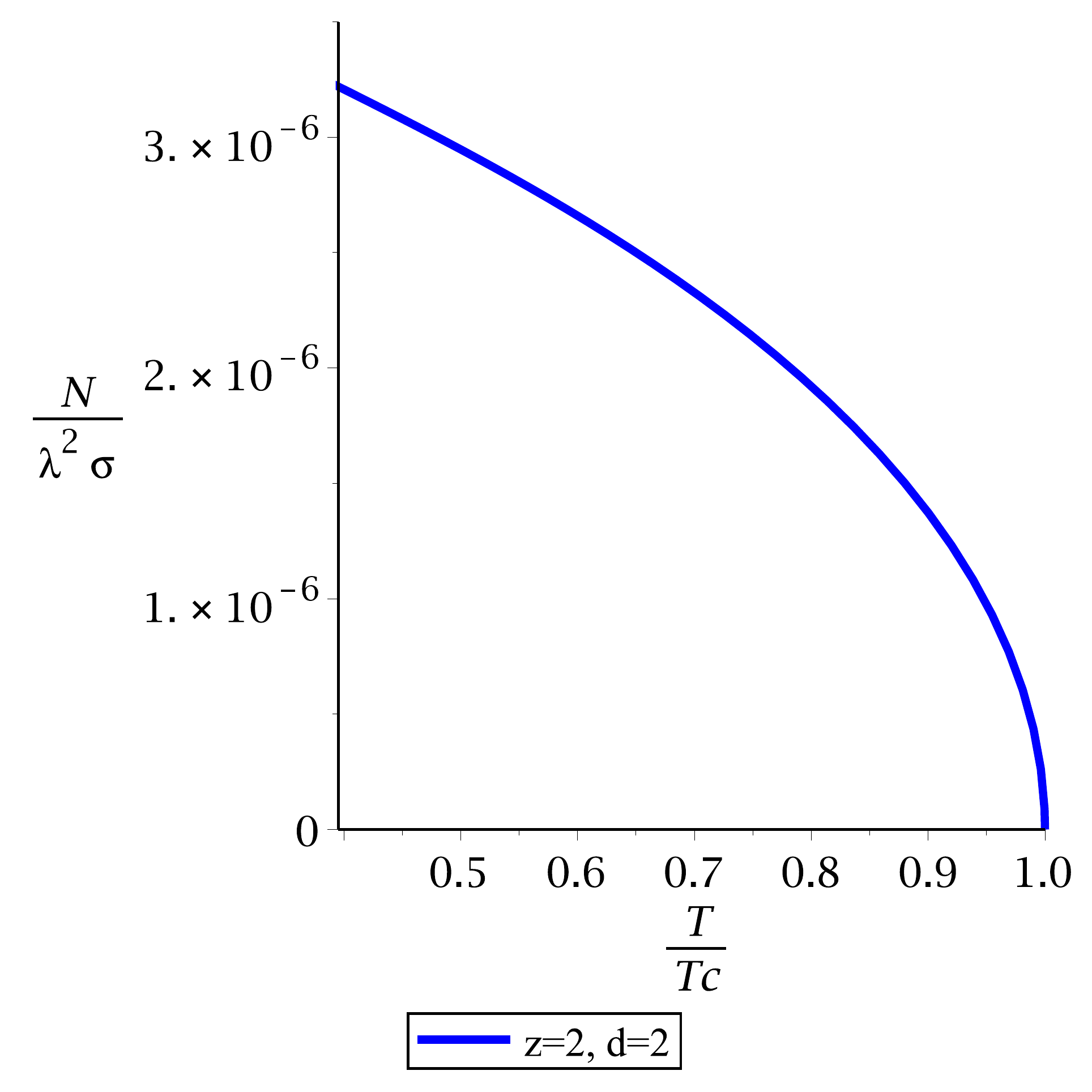}
\includegraphics[width=0.4\textwidth]{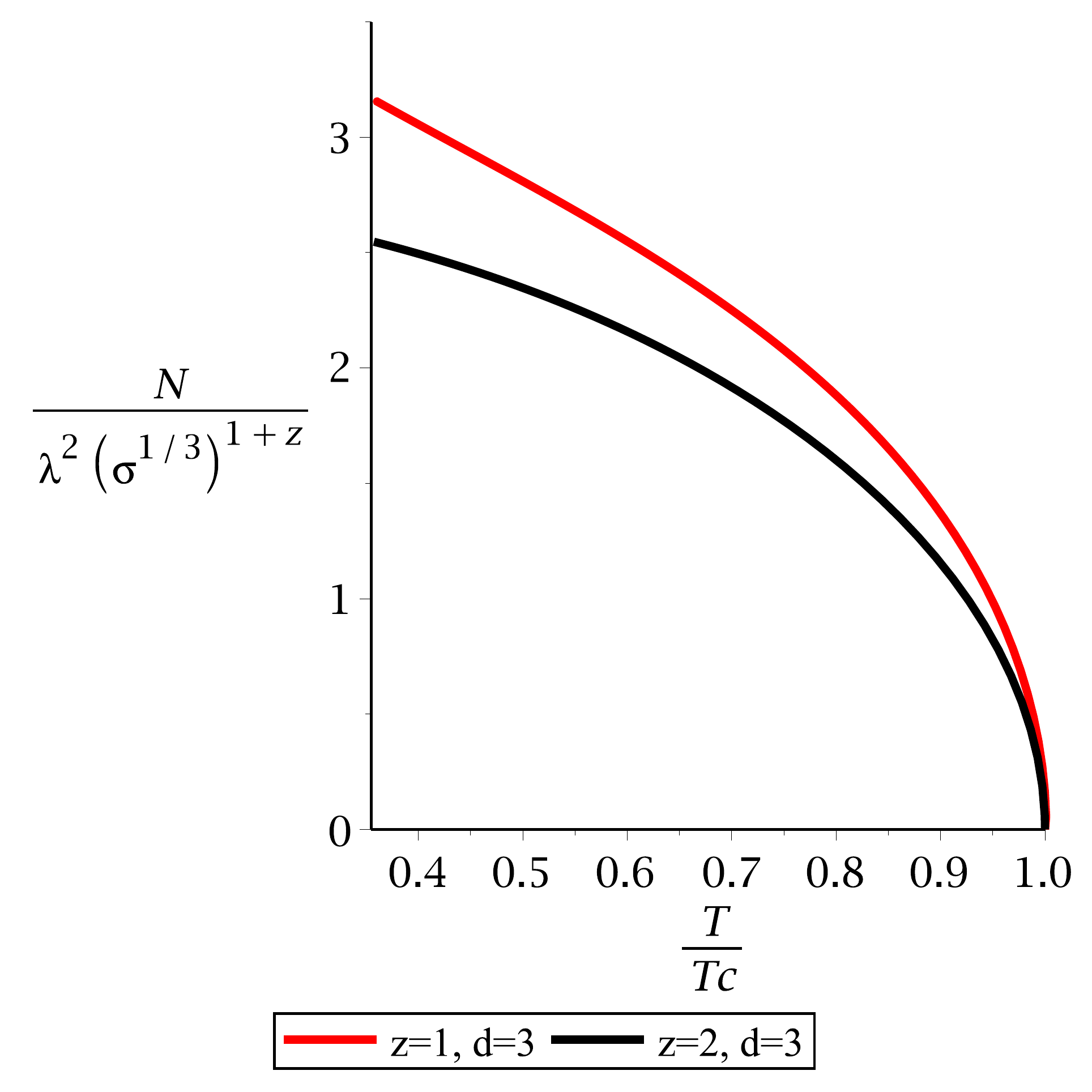}
\includegraphics[width=0.4\textwidth]{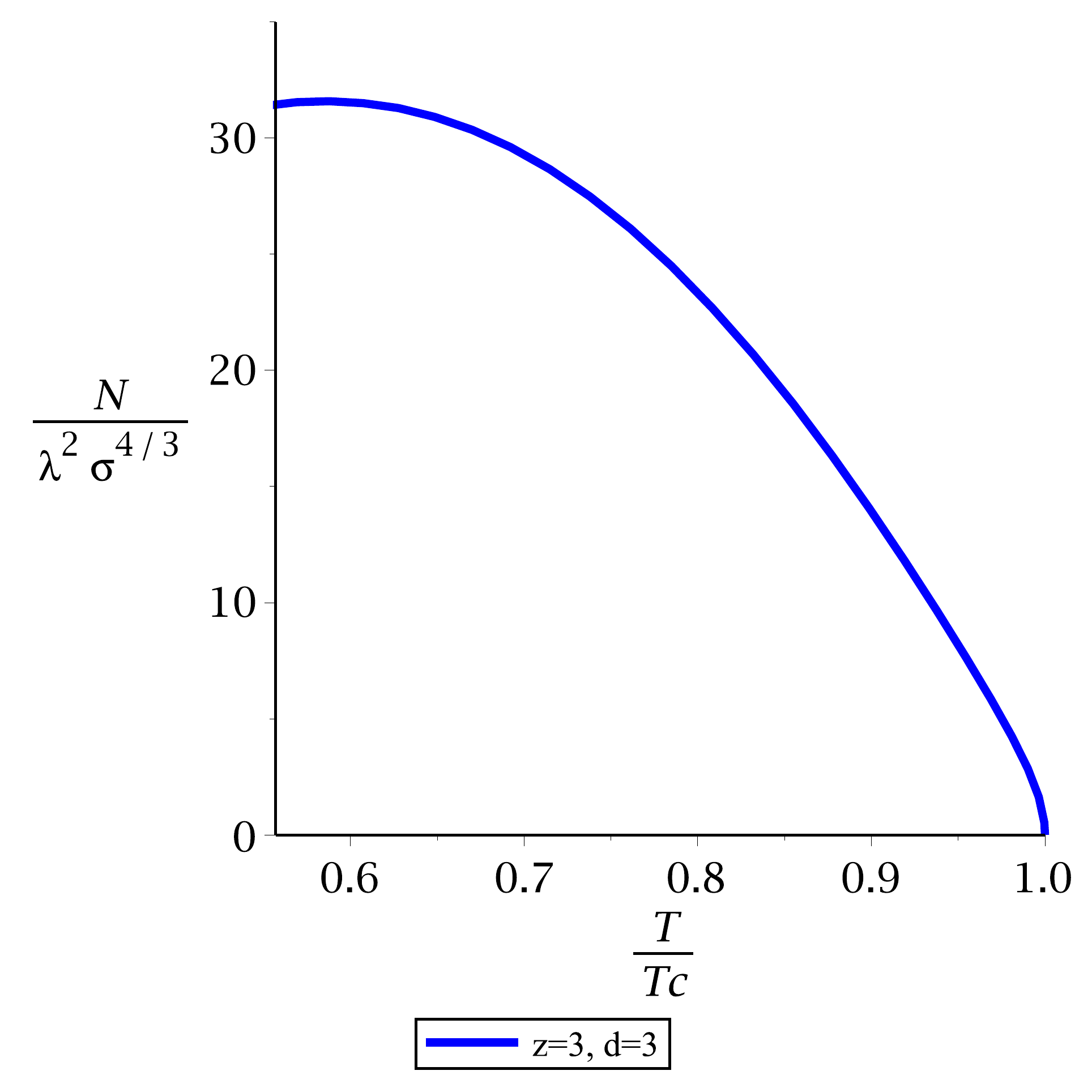}
\caption{The magnetic moment $N$ as a function of temperature.
Top panel: The corresponding temperature $T_{c}/\sqrt{\sigma}\simeq0.6532$ and $T_{c}/\sigma\simeq1.4514$ to $z=1$ and $2$ in the 4D case, respectively. Bottom panel: The corresponding temperature $T_{c}/\sigma^{1/3}\simeq0.6274$, $T_{c}/\sigma^{2/3}\simeq0.6738$ and $T_{c}/\sigma\simeq0.7642$ to $z=1$, $2$ and $3$ in the 5D case, respectively.} \label{FN}
\end{figure}
When the temperature is lower than the critical temperature, we can plot the relationship between $\rho_{+}$ and shooting parameter $\mu$, in order to examine whether $\rho$ gets spontaneous condensation. We find that the solution of source free always appears, which results in the spontaneous magnetization of the system, and breaks the time reversal symmetry in low temperatures. In addition, here it is worth stressing that in the 5D spacetime, the spatial rotational symmetry is also broken spontaneously, since a nonvanishing magnetic moment chooses a direction as a special. When the temperature is lower than the critical one $T_{c}$, we have to solve Eq.~\eqref{eqrhophip} to get the solution of the order parameter $\rho$, and then compute the value of magnetic moment $N$, which is defined in the 4D spacetime by
\begin{equation}\label{eqN}
N=-\lambda^2\int\frac{\rho}{2 r^{3-z}} dr.
\end{equation}
Fig.~\ref{FN} shows the value of magnetic moment $N$ as a function of temperature with various $z$ in 4D and 5D Lifshitz black holes backgrounds. Here to see clearly the effect of the dynamical critical exponent $z$ when we fix the mass squared $m^2$ of the massive 2-form field. We see that when the temperature is lower than $T_{c}$, the non-trivial solution $\rho\neq0$ and spontaneous magnetic moment appears. The numerical results show that this phase transition is a second order one with the behavior $N\propto\sqrt{1-T/T_{c}}$ near the critical temperature for all cases calculated above. The result is still consistent with one in the mean field theory describing the paramagnetism/ferromagnetism phase transition. Note that when taking $z=1$ and $2$ in the 4D and 5D cases, the magnetic moment will decrease with the increasing $z$, especially, the corresponding value of the magnetic moment $z=2$ is smaller than ones to other $z$ in 4D case. However, from the lower right corner of Fig.~\ref{FN}, it easy to see that the magnetic moment increases faster when $T<T_{c}$ than ones of other curves in 5D case. This might be due to the fact that for the cases of $z=2$ in 4D spacetime and $z=3$ in 5D spacetime, there is a logarithmic term in the expansion of the gauge field $\phi$ near the boundary $r\rightarrow\infty$. In addition, at the sufficient low temperature, the backreaction effect of the matter sector on the background geometry becomes important, thus the probe approximation considered here would be no longer valid.

For comparison, we list in Tab.~\ref{tab:tc31} the critical temperature $T_{c}$ and the condensation behavior near $T_{c}$ for the cases of $z=1$, $2$ and $3$ with the fixed squared mass. From the table, we can find that when we increase $z$, $T_{c}$ increases for the fixed D, which indicates that the increasing anisotropy between space and time enhances the phase transition. This can be understood as follows. We can see from Eqs.~\eqref{eqrhophip} that near the horizon, the effective mass of the massive 2-form field decreases as the dynamical critical exponent $z$ increases. This leads to a higher critical temperature as $z$ increases.
\begin{table}
\caption{The critical temperature, magnetic moment and static magnetic susceptibility for the paramagnetic/ferromagnetic phase transition in Lifshitz black hole backgrounds. Here $t=1-T/T_{c}$, the subscript $SL$ denotes the quantity calculated by the semi-analytic method. $N_{c;SL}/\lambda^2 \sigma^{z/d}$, $\lambda^2/\chi_{c;SL} \sigma^{z/d}$, $N/\lambda^2 \sigma^{z/d}$ and $\lambda^2/\chi \sigma^{z/d}$ are calculated near $T_c$. Here each coefficient from $C1$ to $\gamma$ should be multiplied by $10^2$ for the case of $z=2$ in 5D spacetime, but the order of magnitude when we take $z=3$ in 5D spacetime is $10^7$.}
\label{tab:tc31}
\centering
%\begin{tabular}{c c c  c c c  c c c c c c c}
\begin{tabular}{p{0.5cm}p{0.5cm}p{0.5cm}p{1.1cm}p{1.1cm}p{1.1cm}p{1.1cm}p{1.1cm}p{1.7cm}p{1.7cm}p{1.6cm}p{1.6cm}p{1.2cm}}
\toprule[0.7pt]
\hline\hline \midrule[0.7pt]
  $D$& $z$ & $k$ & $C1$ & $N1$ & $a1$ & $a0$ & $\gamma1$ & $\frac{N_{c;SL}}{\lambda^2 \sigma^{z/d}}$ & $\frac{\lambda^2}{\chi_{c;SL} \sigma^{z/d}}$ & $\frac{N}{\lambda^2 \sigma^{z/d}}$ & $\frac{\lambda^2}{\chi \sigma^{z/d}}$ & $T_c/\sigma^{z/d}$  \\ \midrule[0.6pt]
     $4$& $1$ &$3$ &$0.234$ &$1.837$ &$0.477$ &$1.849$ &$1.838$ &$1.341 t^{1/2}$ &$-2.996 t$ & \multirow{2}{*}{$1.332 t^{1/2}$} & \multirow{2}{*}{$-2.966 t$} & \multirow{2}{*}{$0.653$} \\
     $4$& $1$ &$4$ &$0.189$ &$2.042$ &$0.727$ &$2.284$ &$2.042$ &$1.341 t^{1/2}$ &$-2.996 t$ \\
     $4$& $2$ &$3$ &$0.212$ &$5.900$ &$0.588$ &$3.078$ &$6.136$ &$19.218 t^{1/2}$ &$-0.170 t$ & \multirow{2}{*}{$19.27 t^{1/2}$} & \multirow{2}{*}{$-0.57 t$} & \multirow{2}{*}{$1.451$} \\
     $4$& $2$ &$4$ &$0.175$ &$6.496$ &$0.864$ &$3.732$ &$6.757$ &$19.218 t^{1/2}$ &$-0.170 t$ \\
     $5$& $1$ &$3$ &$0.203$ &$5.757$ &$0.504$ &$9.273$ &$5.987$ &$32.913 t^{1/2}$ &$-0.538 t$ & \multirow{2}{*}{$19.09 t^{1/2}$} & \multirow{2}{*}{$-0.508 t$} & \multirow{2}{*}{$0.627$} \\
     $5$& $1$ &$4$ &$0.169$ &$6.325$ &$0.734$ &$11.191$ &$6.577$ &$32.913 t^{1/2}$ &$-0.538 t$ \\
     $5$& $2$ &$3$ &$0.002$ &$0.139$ &$0.007$ &$0.171$ &$0.215$ &$25.468 t^{1/2}$ &$-0.088 t$ & \multirow{2}{*}{$13.692 t^{1/2}$} & \multirow{2}{*}{$-0.05 t$} & \multirow{2}{*}{$0.674$} \\
     $5$& $2$ &$4$ &$0.001$ &$0.149$ &$0.010$ &$0.198$ &$0.231$ &$25.468 t^{1/2}$ &$-0.088 t$ \\
     $5$& $3$ &$3$ &---&--- &---&--- &---&$21.823 t^{1/2}$ &$-0.039 t$ & \multirow{2}{*}{$744.075 t^{1/2}$} & \multirow{2}{*}{$-0.95 t$} & \multirow{2}{*}{$0.764$} \\
     $5$& $3$ &$4$ &---&--- &---&--- &---&$21.823 t^{1/2}$ &$-0.039 t$ \\ \hline \midrule[0.7pt]
\bottomrule[0.7pt]
\end{tabular}
\end{table}

%\subsubsection{Static magnetic susceptibility}
In the following we calculate the static magnetic susceptibility in this probe limit,
defined by
\begin{equation}\label{definechi}
\chi=\lim_{B\rightarrow0}\frac{\partial N}{\partial B}.
\end{equation}
Based on the previous analysis~\cite{Cai:2015bsa}, the magnetic susceptibility is still obtained by solving
\begin{eqnarray}\label{eqchi}
\rho''+(\frac{f'}{f}+\frac{z-1}{r})\rho'-\frac{1}{r^2 f}[m^2+\frac{4 J \mu^2 (2-z)^2}{m^4 r^{4}}]\rho+\frac{B}{r^2 f}
&=&0,~~(z\neq d) \nonumber\\
\rho''+(\frac{f'}{f}+\frac{z-1}{r})\rho'-\frac{1}{r^2 f}[m^2+\frac{4 J \mu^2}{m^4 r^{4}}]\rho+\frac{B}{r^2 f}
&=&0.~~(z=d)
\end{eqnarray}
\begin{figure}
\centering
\includegraphics[width=0.4\textwidth]{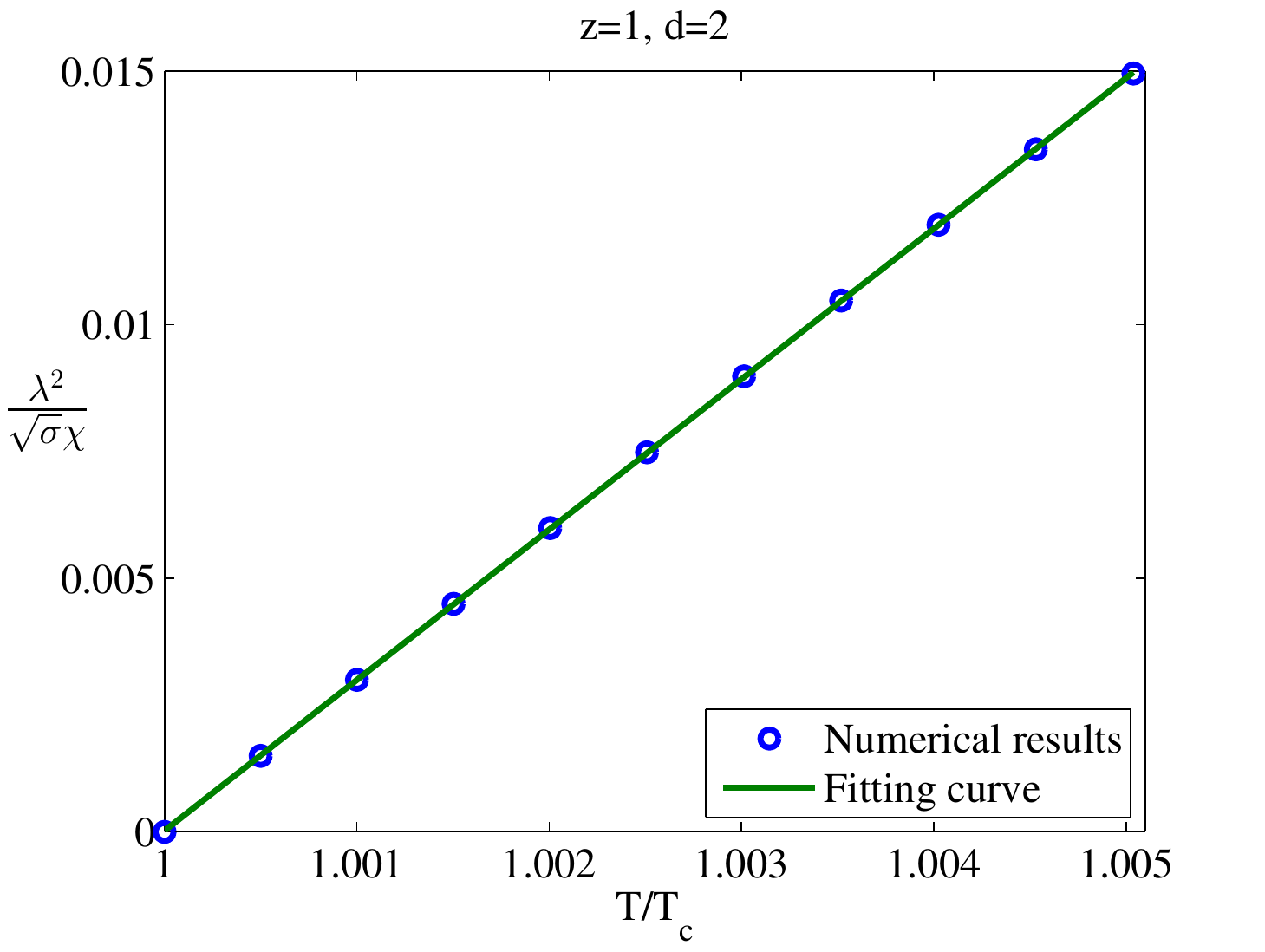}
\includegraphics[width=0.4\textwidth]{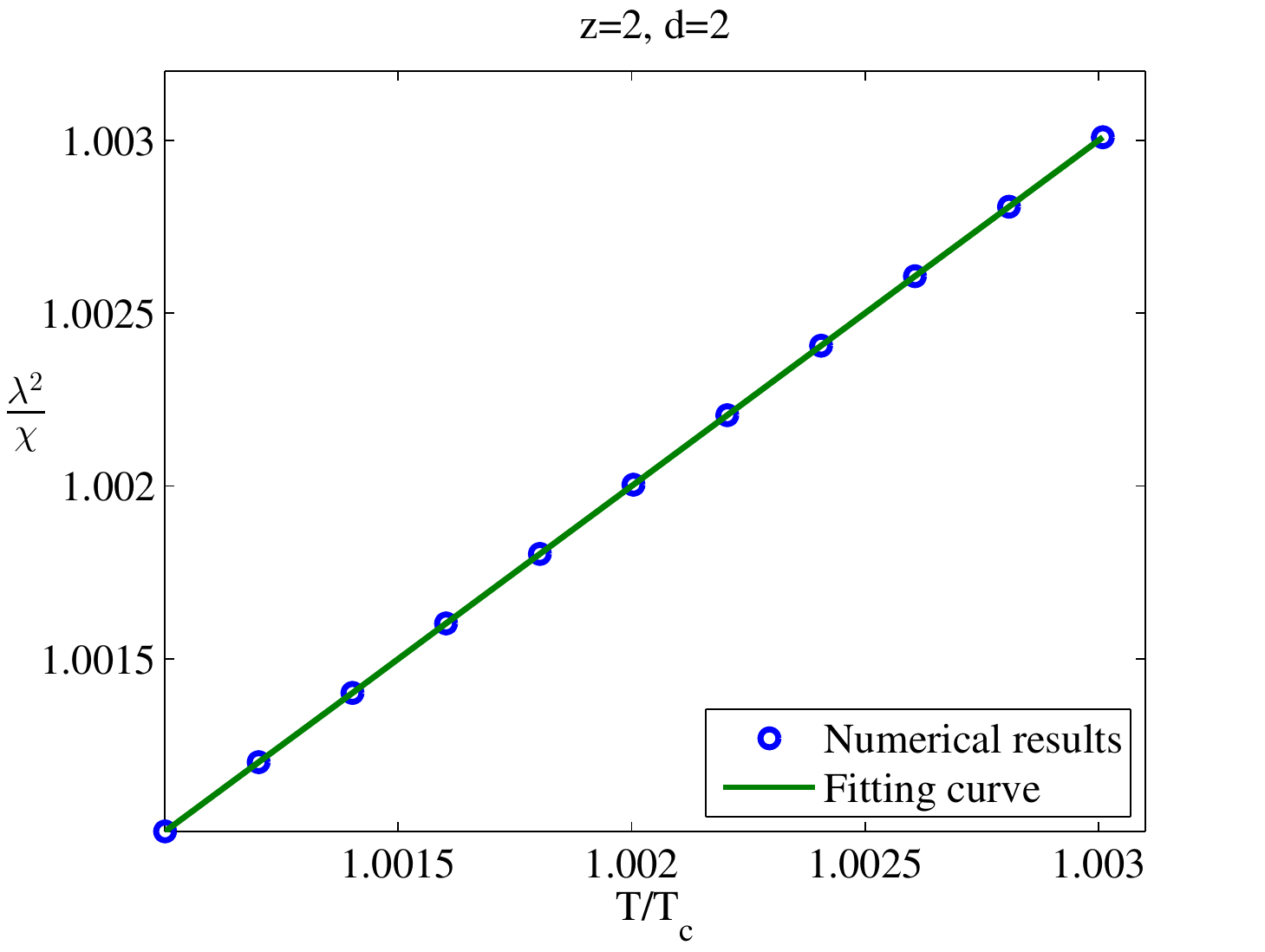}
\includegraphics[width=0.4\textwidth]{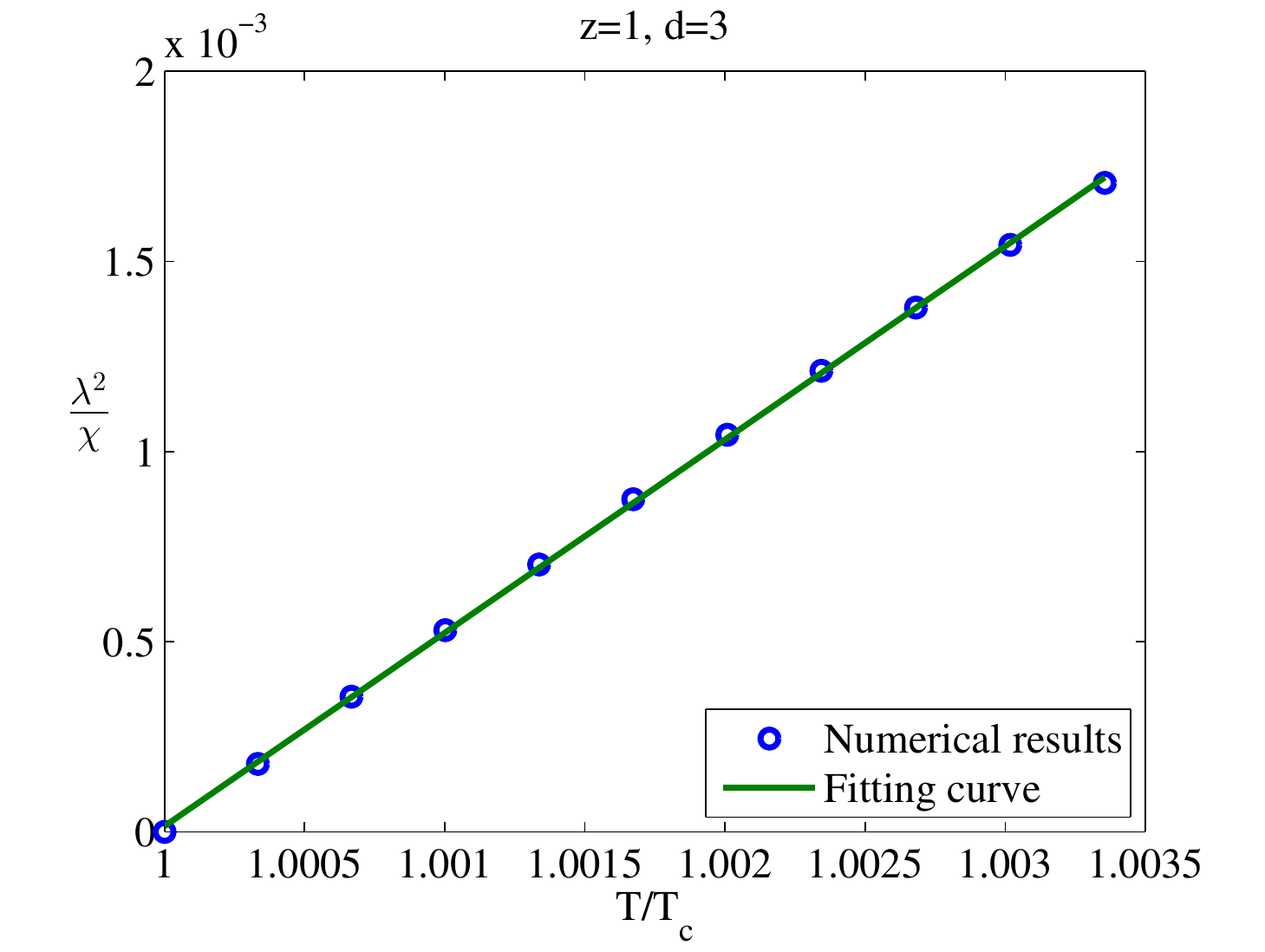}
\includegraphics[width=0.4\textwidth]{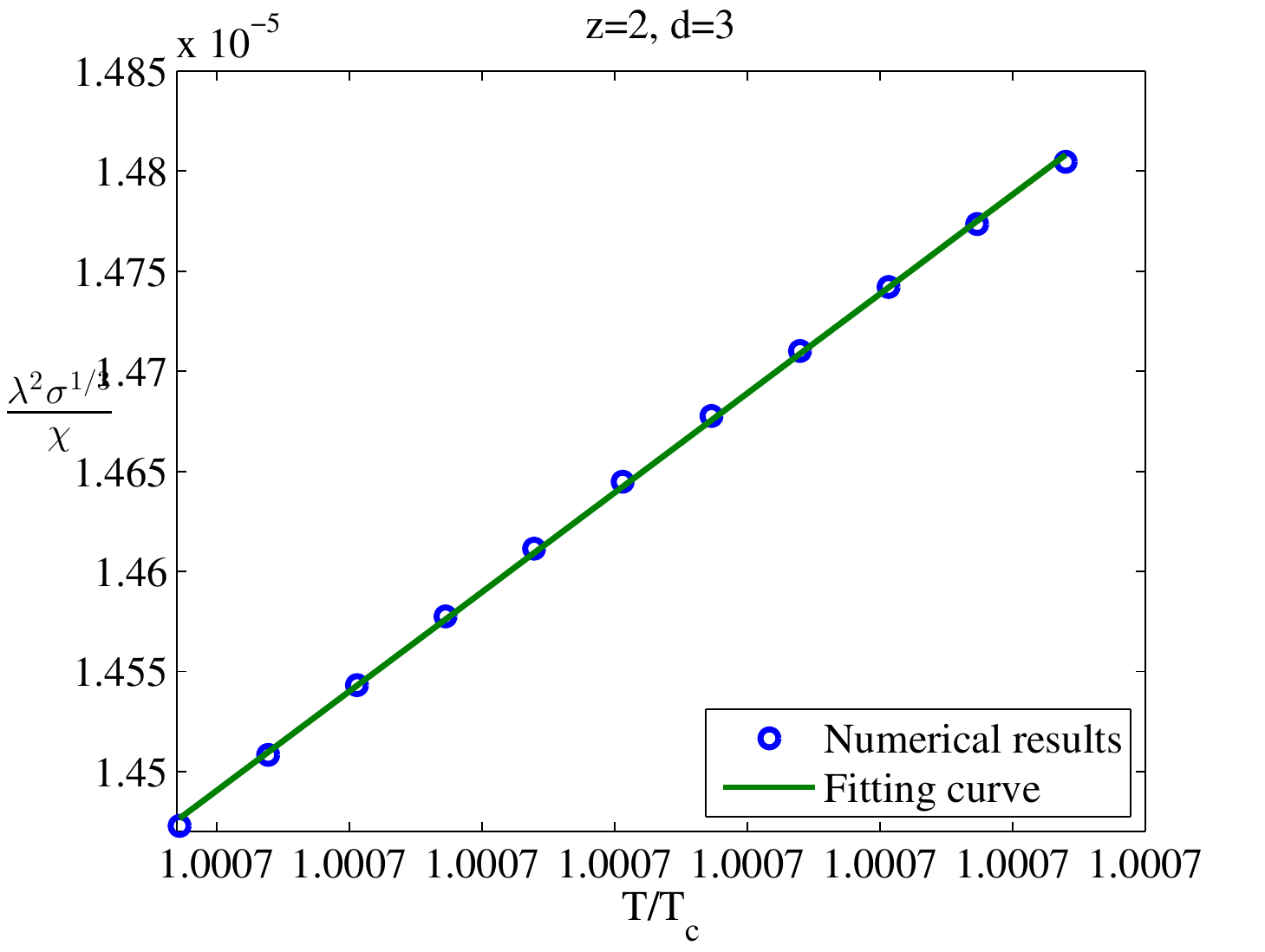}
\includegraphics[width=0.4\textwidth]{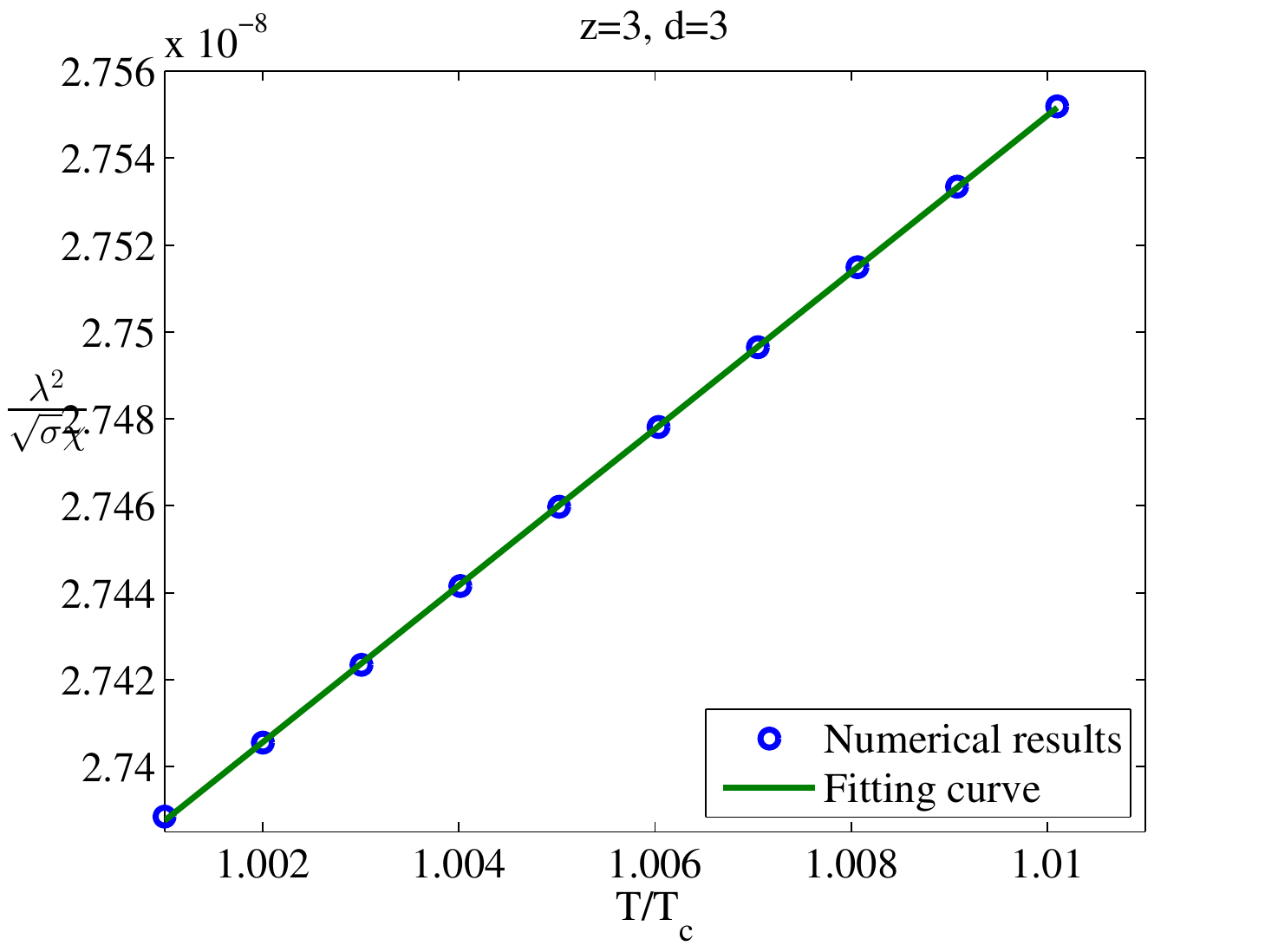}
\caption{The behavior of the inverse susceptibility density in the paramagnetic phase near the critical temperature when $m^2=-J=1/8$. Here we set $2 \kappa^2=1$ for convenience. Top panel: $z=1$ and $2$ in the 4D case from left to right. Bottom panel: $z=1$, $2$ and $3$ in the 5D case from left to right.} \label{Fchi}
\end{figure}
Thus we can also get $\frac{\lambda^{2}}{\chi \sqrt{\sigma}}=2.966 (T/T_{c}-1)$ and $\frac{\lambda^{2}}{\chi\sigma}=0.5751 (T/T_{c}-1)$ when setting the magnetic field $B=1$ and taking $z=1$ and $2$ in the 4D case,
they satisfy the Curie-Weiss law of ferromagnetism in the region of $T>T_{c}$, respectively. This conclusion is similar to the one in the 5D case.
The inverse susceptibility density in paramagnetic is shown in Fig.~\ref{Fchi}.

\subsection{DC conductivity in the ferromagnetic phase}
As we know, the electric transport is also an important property in the materials involving spontaneous magnetization. Now let us study how the DC conductivity is influenced by spontaneous magnetization in this model. In order to simplify our computation in technology, we will work in the probe limit by neglecting back reactions of all the matter fields. This limit can give out the main features near the critical temperature. However, in the case of near zero temperature, we have to consider the model with full back reaction, which will be our work in the near future.

To compute the conductivity, we have to consider some perturbations for gauge field with harmonically time varying electric field. Due to the planar symmetry at the boundary, the conductivity is isotropic. Thus for simplicity, we just compute the conductivity along the $x$-direction. According to the dictionary of AdS/CFT, we consider the perturbation  $\delta A_x=\epsilon a_x(r)e^{-i\omega t}$. In the probe limit, this perturbation will also lead to the perturbations of polarization field in the first order of $\epsilon$. As a result, we have to consider the perturbations for all the components of gauge field and polarization field. However, if we only care the conductivity in the low frequency limit,
i.e., $T\gg\omega\rightarrow0$, the problem can be simplified. In the low frequency limit, we only need turn on the three perturbations,
\begin{equation}\label{pert1}
\begin{split}
\delta A_x&=\epsilon a_x(r)e^{-i\omega t},\\
M_{rx}&=\epsilon C_{rx}(r)e^{-i\omega t},\\
M_{ty}&=\epsilon C_{ty}(r)e^{-i\omega t},
\end{split}
\end{equation}
and  corresponding equations to the three perturbations in the low frequency limit with Lifshitz scaling $z$ read
\begin{subequations}\label{pert2}
\begin{align}
C_{ty}''+(\frac{1-z}{r})C_{ty}'-\frac{m^2C_{ty}}{r^2f}-\frac{4Jp\rho C_{rx}}{r^2}+O(\omega)=0,\label{pert2a}\\
C_{rx}-\frac{a_x'}{m^2}-\frac{4Jp\rho C_{ty}}{r^{2z+2}f m^2}+O(\omega)=0,\label{pert2b}\\
[r^{z+1}f(a_x'-\lambda^2C_{rx}/4)]'+\frac{a_x\omega^2}{r^{z+1}f}+O(\omega)=0,\label{pert2c}
\end{align}
\end{subequations}
with $p$ and $\rho$ determined by Eqs.~\eqref{eqrhophip}. Here $O(\omega)$ is the terms with order of $\omega$ which can be neglected when $\omega\rightarrow0$. In general, the term  $\omega^2/r^{z+1}f(r)$ can not be neglected since $f(r)$ is zero at the horizon, which makes the limit of $\omega\rightarrow0$ ambiguous. However, at the horizon, if we impose the ingoing conditions for $C_{rx}, C_{ty}$ and $a_x$,
\begin{equation}\label{init4}
\begin{aligned}
C_{ty}&=e^{-i\omega r_*}[C_{ty}^{(0)}+C_{ty}^{(1)}(r-r_h)+\cdots],\\
C_{rx}&=e^{-i\omega r_*}[C_{rx}^{(0)}+C_{rx}^{(1)}(r-r_h)+\cdots], \\
a_x&=e^{-i\omega r_*}[a_{x}^{(0)}+a_{x}^{(1)}(r-r_h)+\cdots]
\end{aligned}
\end{equation}
with $r_*=\int dr/(r^{z+1}f)$,  we find the system has a well-defined limit when $\omega\rightarrow0$ if $T\neq0$.
At the AdS boundary with the source free condition, we have the following asymptotic solutions,
\begin{equation}\label{pert3}
\begin{split}
C_{ty}&=C_{ty+}r^{(z+\delta)/2}+C_{ty-}r^{(z-\delta)/2}+\cdots,\\
C_{rx}&=-\frac{za_{x-}}{r^{z+1} m^2}+\cdots,~a_x=a_{x+}+\frac{a_{x-}}{r^{z}}+\cdots.
\end{split}
\end{equation}
Here $\delta=\sqrt{4m^2+z^2}$. Then the gauge/gravity duality implies  that electric current $\langle J_x\rangle=a_{x-}$ and the DC conductivity is given by
\begin{equation}\label{DCconduc}
\sigma=\lim_{\omega\rightarrow0}\frac{a_{x-}}{i\omega a_{x+}}.
\end{equation}
As a holographic application of the membrane paradigm of black holes, we can directly obtain the DC conductivity from  Eq.~\eqref{pert2} using the method proposed by Iqbal and Liu in~\cite{Iqbal:2008by}. In fact, the transport coefficients in the dual field theory can be obtained from the horizon geometry of the dual gravity in the low frequency limit. Applying this into $U(1)$ gauge field, this conclusion implies that the DC conductivity is given by the coefficient of the gauge field kinetic term evaluated at the horizon. To see this, we assume that $T>0$ and $\omega\rightarrow0$, then we can neglect all the terms of $\omega$ in Eqs.~\eqref{pert2}. We first note that,
\begin{equation}\label{pert4}
\begin{split}
&\lim_{r\rightarrow\infty}r^{z+1} f(r)(a_x'-\lambda^2C_{rx}/4)=r^{z+1}z \left(\frac{-\langle J_x\rangle}{r^{z+1}}+ \frac{\lambda^2\langle J_x\rangle}{4r^{z+1}m^2}\right)\\
&=-z(1-\lambda^2/4m^2)\langle J_x\rangle.
\end{split}
\end{equation}
 Eq.~\eqref{pert2c} shows that this quantity is conserved along the direction $r$. So at the horizon, using Eqs.~\eqref{pert2b} and \eqref{pert2c}, we have,
\begin{equation}\label{pert5}
\begin{aligned}
&-z(1-\lambda^2/4m^2)\langle J_x\rangle=\lim_{r\rightarrow1}\left.r^{z+1}f\left(a_x'-\lambda^2C_{rx}/4\right)\right|_{r=r_{+}},\\
&=r^{z+1}f\left[\left(1-\frac{\lambda^2}{4m^2}\right) a_x'-\frac{\lambda^{2}Jp\rho C_{ty}}{m^2f r^{2z+2}}\right]_{r=r_{h}}.
\end{aligned}
\end{equation}
Combining Eqs.~\eqref{pert2a} with~\eqref{pert2b} and considering the fact that $C_{ty}$ is regular at the horizon, we have,
\begin{equation}\label{pert6}
\left[m^2+\frac{16J^2p^2\rho^2}{m^2r^{2z+2}}\right]C_{ty}=- \frac{4Jp\rho}{m^2}fa_x'+(z-1)fC_{ty}'
\end{equation}
at $r\rightarrow r_{h}^+$. Thus we have  from Eqs.~\eqref{pert6} and~\eqref{pert5} that
\begin{equation}\label{pertb1}
\begin{aligned}
&-z(1-\lambda^2/4m^2)\langle J_x\rangle\\
&=\lim_{r\rightarrow r_{h}^{+}}r^2fa_x'(1-\frac{\lambda^{2}}{4 m^2})\left[1+\frac{4J^2p^2\rho^2\lambda^2}{(m^2-\frac{\lambda^2}{4})(m^4+16J^2p^2\rho^2/r^{2z+2})}\right].
\end{aligned}
\end{equation}
Now let us take the ingoing condition for $a_x$ at the horizon, which tells us that,
\begin{equation}\label{ingoing1}
r^{z+1}fa_x'=\frac{d}{dr_*}a_x=-i\omega a_x,~~\text{at}~r\rightarrow r_{h}^+,
\end{equation}
finally we get,
\begin{equation}\label{pertb2}
\langle J_x\rangle=\frac{i\omega a_x(r_{+})}{z}\left[1+\frac{4J^2p_0^2\rho_0^2\lambda^2}{(m^2-\frac{\lambda^2}{4})(m^4+16J^2p_0^2\rho_0^2/r_{+}^{2z+2})}\right].
\end{equation}
Here $p_0$ and $\rho_0$ are the initial values of $p(r)$ and $\rho(r)$ at the horizon, which can be computed from Eq.~\eqref{eqrhophip}. In the low frequency limit, Eq.~\eqref{pert2} imply that the electric field is constant, i.e., $\lim_{r=r_{h}}a_x(r)=a_{x +}$. It follows that we can obtain the DC conductivity near the horizon as,
\begin{equation}\label{DCexp}
\sigma=\frac{1}{z}\left[1+\frac{4Jp_0^2\rho_0^2\lambda^2}{(m^2-\frac{\lambda^2}{4})(m^4+16J^2p_0^2\rho_0^2/r_{h}^{2z+2})}\right].
\end{equation}
\begin{figure}
\begin{center}
\includegraphics[width=0.33\textwidth]{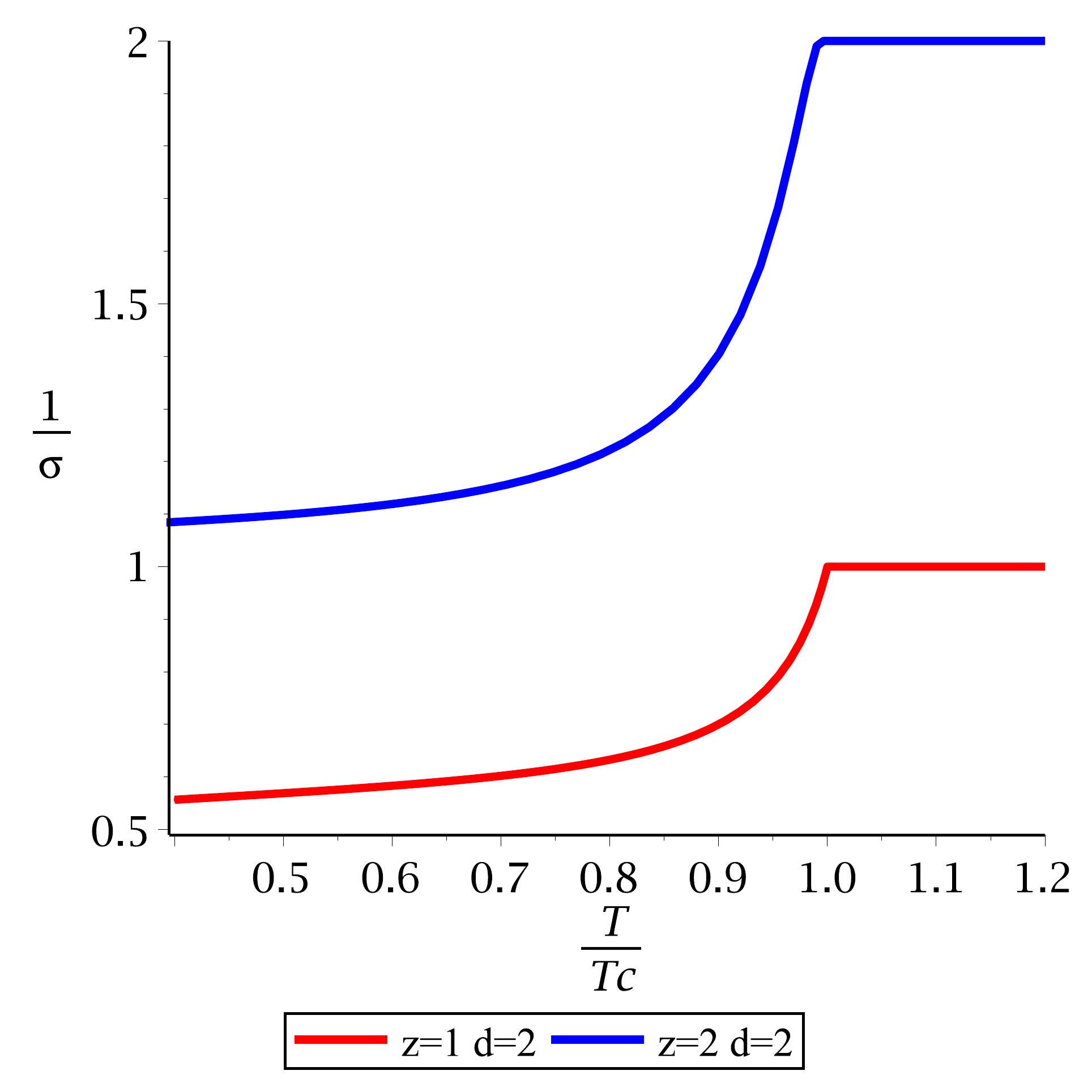}
\includegraphics[width=0.33\textwidth]{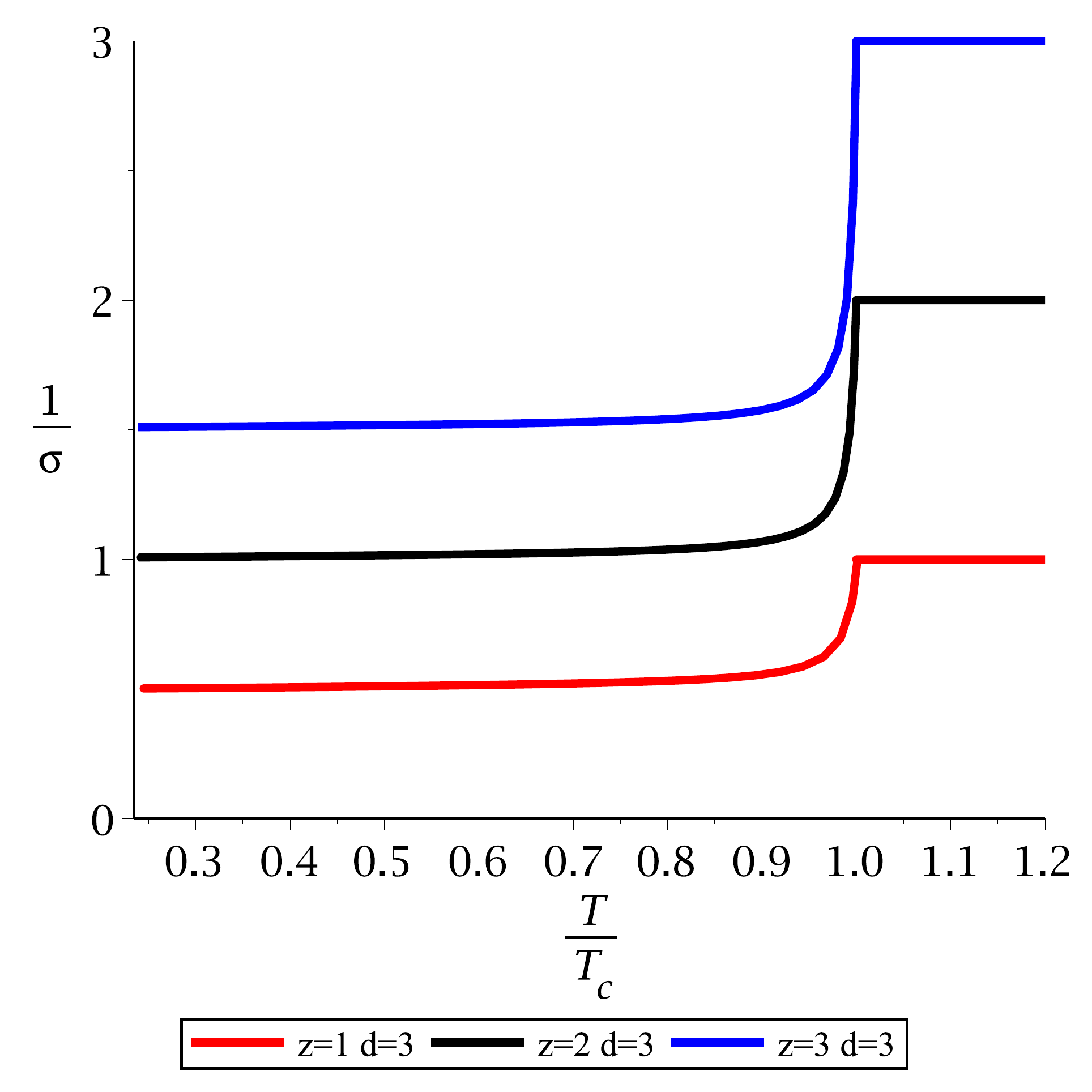}
\includegraphics[width=0.3\textwidth]{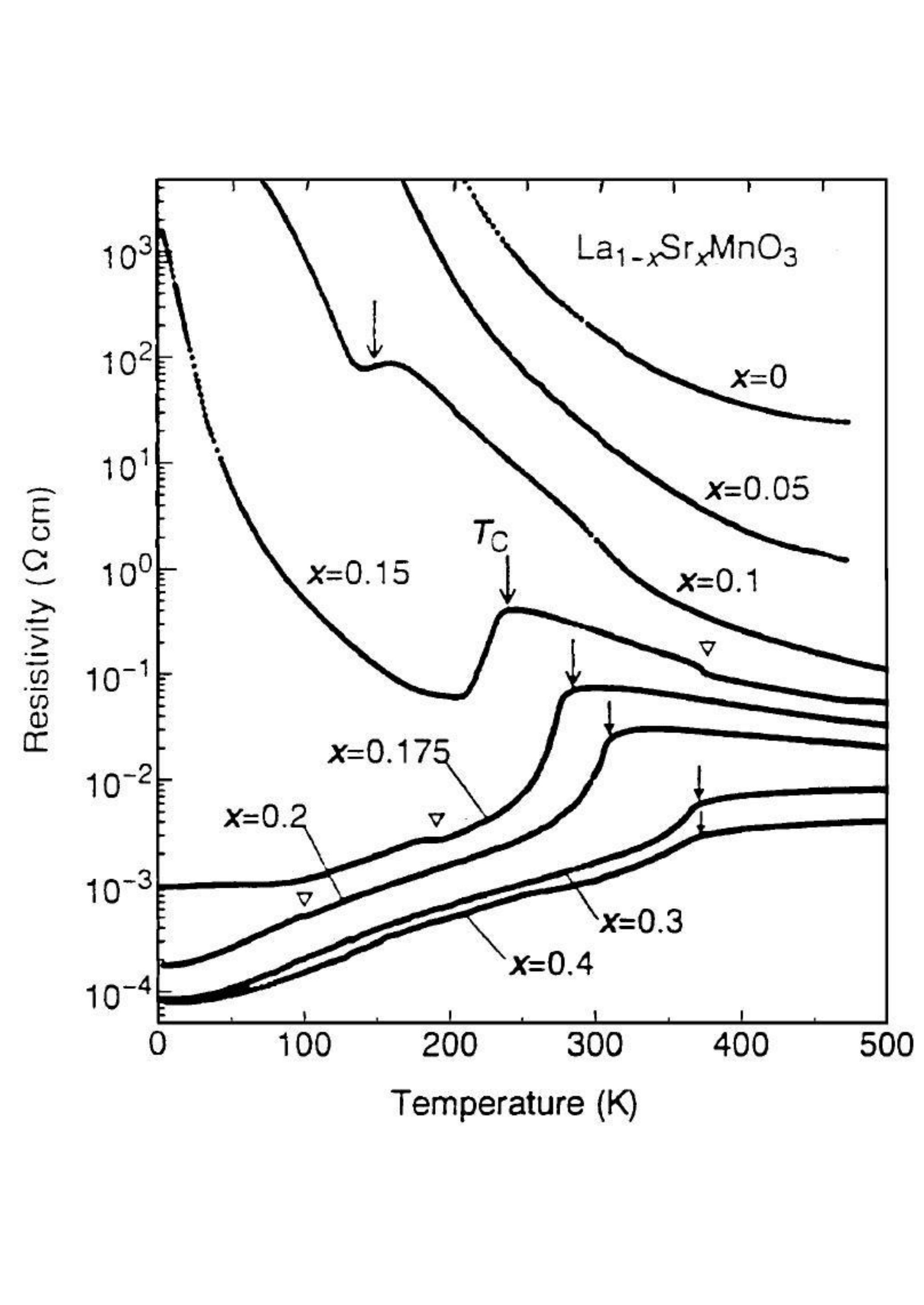}
\caption{Left panel: DC resistivity vs temperature in our model. Here we choose parameters as $m^2=-J=1/8$ and $\lambda=1/2$. Right panel: Temperature dependence of resistivity for various single crystals of La$_{1-x}$Sr$_x$MnO$_3$. Arrows indicate the Curie temperature. For more details, see Ref.~\cite{A.Y.T}. }
\label{TO1}
\end{center}
\end{figure}
With the appearance of ferromagnetism, DC resistivity decreases when the sample gets cooling, which shows in many interesting phenomena in condensed matter physics, especially in a class of manganese oxides which are widespread because of the discovery of colossal magnetoresistance (CMR)~\cite{E.T.A,E.L.N}. Note that this effect has a complete different physical origin from the ``giant" magnetoresistance observed in layered and clustered compounds. In recent twenty years, CMR is among the main topics of study within the area of strongly correlated electron systems and its popularity is reaching the level comparable to that of the high-temperature superconducting cuprates. Here the expression of ~\eqref{DCexp} is just appropriate for the case of 4D spacetime. However, for the 5D Lifshitz spacetime, we only need to replace the coefficient $1/z$ in front of the bracket in~\eqref{DCexp} with $1/(z+1)$. It is no difficult to find that the dynamical exponent $z$ has no effect on the shape of curve for the fixed D from Fig.~\ref{TO1}, But it affects the value of DC resistivity when the sample gets cooling, i.e., the bigger the value of $z$, the bigger the DC resistivity although it decreases with the lower temperature. Moreover, the DC resistivity decreases faster near the critical temperature $T_c$ in the case of 5D than the one in the case of 4D. In the right panel of Fig.~\ref{TO1}, we show the experimental data from a typical CMR material La$_{1-x}$Sr$_x$MnO$_3$ as an example. We see that our model gives a very similar behavior to the latter in a composition range of $x\geq0.175$. Meanwhile Fig.~\ref{TO1} shows that the effect of dynamical exponent $z$ on DC resistivity is different from the composition $x$, i.e., the increase of the composition $x$ will result in the decrease of DC resistivity. Of course, we should mention here that there still exist some differences between our model result and experimental data on CMR. In general, when $T>T_c$, the material shows a semiconductor or insulator behavior and the DC resistivity increases with cooling the sample which has been realized by introducing a massive 2-form field and a dilaton field coupled with U(1) gauge field in asymptotic AdS black brane background~\cite{Cai:2015bsa}. In our model, however, the DC resistivity is a constant when $T>T_c$, which is similar to the Ref.~\cite{Cai:2015jta}. So this model only gives partial property of CMR when $T< T_c$. But this is an exciting and enlightening result, because it implies that this model still can lead to a possibility to build a holographic CMR model in Lifshitz black hole and to investigate this typical and important strong correlated electrons system in the AdS/CFT setup.

\section{Semi-analytic calculations near the critical temperature}

In this section, to complement the numerical calculations, we study magnetic moment and static magnetic susceptibility by using the semi-analytic method, which is different from the analytic method in holographic superconductors but it seems more accurate. Now we focus on the case of 4-dimensional Lifshitz black hole. It is convenient to make a coordinate transformation by $u=r_h/r$. As $p$ can be solved directly, then we put it into the equation of $\rho(r)$, and get when $z\neq d$
\begin{equation}\label{Eoms1}
\begin{split}
\rho''+\left(\frac{3-z}{u}+\frac{f'}{f}\right)\rho'-[\frac{m^2}{u^2 f}+\frac{4 J \mu^2u^2(2-z)^2}{(m^2-4 J\rho^2u^4)^2f}]\rho&+\frac{B}{u^2 f}=0.
\end{split}
\end{equation}
As we will care about the behavior of $T\rightarrow T_c$, the value of $\rho$ will be a small quantity near the transition point. In this case, we can make a Taylor's expansion on the nonlinear term of $\rho$ in Eq.~\eqref{Eoms1} as,
\begin{equation}\label{Taylor1}
\frac{4 J \mu^2u^2(2-z)^2}{(m^2-4 J\rho^2u^4)^2}=\frac{4 J \mu^2u^2(2-z)^2}{m^4}+\frac{32J^2\mu^2\rho^2u^6(2-z)^6}{m^6}+\mathcal{O}(\rho^4)
\end{equation}
Note that for the case of $z=d$, all the terms about $(2-z)$ in molecules will be replaced by one, and some expressions in the following will also be changed. Here we don't talk about it in details. When neglecting the high order terms,  Eq.~\eqref{Eoms1} can be rewritten as
\begin{equation}\label{rhop}
\begin{split}
&\widehat{L}\rho=\widetilde{J}_f\rho^3u^{9-z}+Bu^{1-z},\\
&\widehat{L}=-\frac{d}{dz}\left[u^{3-z}f(u)\frac{d}{du}\right]+q(u),\\
&q(u)=m^2u^{1-z}+\frac{4 J \mu^2u^{5-z}(2-z)^2}{m^4},\\
&\widetilde{J}_f=-32J^2\mu^2(2-z)^6/m^6<0.
\end{split}
\end{equation}
Up to the order of $\rho^4$, the part of polarization field in action  \eqref{1} can be written as,
\begin{equation}
\begin{split}
\frac{S(T,B;\rho)}{\lambda^2V_2}&=\left.(\frac{u^{3-z}}2f\rho'\rho+u^{2-z}f\rho^2)\right|^{z_h}_{0}\\
%&=\int_{z_h}^{0} dz\left[-\frac{d}{dz}(z^2\frac12f\rho'\rho+zf\rho^2)+\frac {\rho}2\widehat{L}\rho+B\rho-\frac{\widetilde{J}_f}4z^8\rho^4\right]\\
&-\int_{0}^{u_h} dz\left[\frac {\rho}2\widehat{L}\rho+B\rho u^{1-z}-\frac{\widetilde{J}_f}4u^{9-z}\rho^4\right],
\end{split}
\end{equation}
which is a function of $T$ and $B$, but a functional of $\rho$.
The asymptotic solution for~\eqref{rhop} is
\begin{equation}\label{symp1}
\rho=\widetilde{\rho}+\frac{B}{m^2},~\text{with}~\widetilde{\rho}=\rho_+u^{\Delta_{-}}+\rho_-u^{\Delta_{+}}.
\end{equation}
The source free condition is  $\rho_+=0$ as $u\rightarrow0^+$. Under this, the grand thermodynamic potential or free energy in grand canonical ensemble $\Omega$ is,
\begin{equation}\label{action1}
\begin{split}
\Omega(T,B;\rho)&=\widetilde{\Omega}(T,B;\rho)V_2\\
&=\lambda^2V_2\int_{0}^{u_h} dz\left[\frac {\rho}2\widehat{L}\rho+B\rho u^{1-z}-\frac{\widetilde{J}_f}4u^{9-z}\rho^4\right].
\end{split}
\end{equation}
According to thermodynamic relationship,
\begin{equation}\label{thero1}
\begin{split}
&d\Omega(T,B)=-SdT-V_2NdB\\
&\Rightarrow N/\lambda^2=-\frac1V_2\left(\frac{\partial \Omega(T,B)}{\partial B}\right)_{T}.
\end{split}
\end{equation}
It seems that the magnetic moment should be,
\begin{equation}\label{defN0}
N=-\frac{\lambda^2}{V_2}\left(\frac{\partial \Omega(T,B;\rho)}{\partial B}\right)_{T,\rho}=-\lambda^2\int_{0}^{u_h} \rho u^{1-z} du.
\end{equation}
However, comparing this result with the previous definition of the magnetic moment,
\begin{equation}\label{defN2}
N/\lambda^2=-\int_{0}^{u_h} \frac{\rho u^{1-z}}{2} du.
\end{equation}
We find the difference factor $1/2$ between~\eqref{defN0} and~\eqref{defN2}. It should be shown that the expression~\eqref{defN0} is not true. The reason has been explained in~\cite{Cai:2015jta} according to the Euler homogenous function theorem and the scaling transformation. Therefore we can get the definition~\eqref{defN2} and still use it in the following. The key step for computing the grand thermodynamic potential is to structure the Sturm-Liouville problem,\footnote{The method is similar to the one used in Ref.~\cite{Yin:2013fwa}, but is completely different from the Sturm-Liouville (SL) eigenvalue method in Ref.~\cite{Siopsis:2010uq}, there the precision depends on the trial function one chooses.}
which is the following ODE:
\begin{equation}\label{SL1}
\begin{split}
\widehat{P}\rho_n&=\frac{\widehat{L}\rho_n}{\omega(u)}\\
&=\frac1{\omega(u)}\left\{-\frac{d}{du}\left[u^{3-z}f(u)\frac{d}{du}\right]+q(u)\right\}\rho_n=\lambda_n\rho_n,
\end{split}
\end{equation}
with the boundary conditions: one is $|\rho_n(u_h)|$ is required to be finite at $u=u_h$, $f(u_h)=0$, and the other is $\rho_n(0)=0$ at $u\rightarrow0^+$. The weight function $\omega(u)$ can be an arbitrary  positive continuous function in the region of $[0,u_h]$. From a practical point of view, we choose weight function such that the values of $\lambda_n$ will not influence the asymptotic solutions of equation~\eqref{SL1}. There are many choices for weight function. Here we choose $\omega(r)=u^k$ with an integer $k>2$.

when $r\rightarrow\infty$, we note that the asymptotic solution for equation~\eqref{SL1} is,
\begin{equation}\label{asym1}
\rho_n=\rho_+u^{\Delta_{-}}+\rho_-u^{\Delta_{+}}.
\end{equation}
One can find that the second boundary condition corresponds to $\rho_+=0$. Let $\mathcal{L}^2([0,u_h],\omega(u),du)$ be the Hilbert space of square integrable functions on $[0,u_h]$, i.e.,
\begin{equation}\label{H1}
\begin{split}
&\mathcal{L}^2([0,u_h],\omega(u),du)\\
=&\left\{h: [0, u_h]\mapsto\mathbb{R}\left| \int_{0}^{u_h}\omega(u)|h(u)|^2du<\infty\right.\right\}
\end{split}
\end{equation}
with the inner product
\begin{equation}\label{inprod}
\langle h_1,h_2\rangle=\int_{0}^{u_h}\omega(u)h_1(u)h_2(u)du,
\end{equation}
and $D$ be the subspace of $\mathcal{L}^2([0,u_h],\omega(u),du)$ that satisfies the both of boundary conditions, i.e.,
\begin{equation}\label{D1}
\begin{split}
D=&\left\{\forall h\in\mathcal{L}^2([0,u_h],\omega(u),du)\left|h\in C^2[0,u_h]\right.,\right.\\
&\left.h(0)=0,~|h(u_h)|<\infty\right\}.
\end{split}
\end{equation}
Then we can prove that $\widehat{P}$ is the self-adjoint operator on $D$, i.e.,
\begin{equation}\label{slad}
\forall h_1,h_2\in D,\langle h_1,\widehat{P}h_2\rangle=\langle \widehat{P}h_1,h_2\rangle.
\end{equation}
According to the properties of SL problem, the solutions of~\eqref{SL1} form a function basis on $D$  with which one can expand any functions belonging to $D$, i.e.,
\begin{equation}\label{expand0}
\langle\rho_n,\rho_k\rangle=\delta_{nk},
\end{equation}
and
\begin{equation}\label{expand0b}
%\begin{split}
\forall h\in D, \exists\{c_n\}\subset\mathbb{R}, h(u)=\sum_{n=1}^\infty c_n\rho_n(u)
%\end{split}
\end{equation}
with $c_n=\langle\rho_n,h\rangle$.

%\subsection{Compute the grand thermodynamic potential}
Let us now turn our attention to the free energy~\eqref{action1}. For convenience, we will use scaling transformation to set $u_h=1$ in the process of computation, and then transform into the case of fixing charge density in the final results.

Let $\widetilde{\rho}(r)=\rho(r)-B/m^2$ be any function configuration belonging to $D$, in which $\rho(r)$ dose't need to be the solution of EoM~\eqref{rhop}. We can use the eigenfunction $\rho_n$ to expand  $\widetilde{\rho}(r)$ and magnetic moment as,
\begin{eqnarray}\label{exprho1}
\widetilde{\rho}&=&\sum_{n=1}^\infty c_n\rho_n\Leftrightarrow\rho=\sum_{n=1}^\infty c_n\rho_n+\frac{B}{m^2},\\
N&=&-\int_{0}^{1}\frac{\lambda^2Bu^{1-z}}{2m^2}du-\lambda^2\int_{0}^{1}\frac{\widetilde{\rho}u^{1-z}}{2}du=-\int_{0}^{1}\frac{\lambda^2Bu^{1-z}}{2m^2}du-\frac{\lambda^2}2\sum_{n=1}^\infty c_n N_n,
\end{eqnarray}
where $c_n$ and $N_n$ are coefficients,  defined as,
\begin{equation}\label{coeffN}
c_n=\int_{0}^{1}\omega\widetilde{\rho}\rho_ndu,~N_n=\int_{0}^{1}\rho_nu^{1-z}du.
\end{equation}
Let us consider the case of spontaneous magnetization, i.e., the case with $B=0$. In this case, we have
\begin{equation}\label{action2}
%\begin{split}
\widetilde{\Omega}(T,c_n)=\lambda^2\int_{0}^{1} du\left[\frac {\omega\rho}2\widehat{P}\rho-\widetilde{J}_fu^{9-z}\rho^4/4\right],
\end{equation}
with $\rho=\sum_{n=1}^\infty c_n\rho_n$. Using the orthogonal relationship, we have,
\begin{equation}\label{expand2}
\begin{split}
\widetilde{\Omega}(T,c_n)&=\frac{\lambda^2}2\langle\rho, \widehat{P}\rho\rangle-\frac{\lambda^2\widetilde{J}_f}4\int_{0}^{1}u^{9-z}\rho^4du\\
&=\frac{\lambda^2}2\sum_{n=1}^\infty \lambda_nc_n^2-\frac{\lambda^2\widetilde{J}_f}4\int_{0}^{1}u^{9-z}\rho^4du.
\end{split}
\end{equation}
It follows that the nonzero solution appears only when $\lambda_1<0$, i.e., $T<T_c$. Because of $J<0$, we can find that $\widetilde{\Omega}(T, c_n)\geq0$. The minimization of $\Omega(T, c_n)=0$ is achieved only when $c_n=0$, i.e., $\rho=0$.

When $T\rightarrow T_c^-$, we can set $\lambda_1=a_0(T/T_c-1)$ with $a_0>0$ and assume that the off-shell solution is dominated by the first term in \eqref{exprho1} only, i.e.,  $|c_1|\gg c_n$ for $n\geq2$ in \eqref{coeffN}. As  a result,  we have,
\begin{equation}\label{expand3}
\begin{split}
\lambda^{-2}\widetilde{\Omega}(T,c_n)&\simeq \frac12\lambda_1c_1^2-\frac{\widetilde{J}_fc_1^4}4\int_{0}^{1} dz\rho_1^4u^{9-z},\\
&\simeq \frac12a_0(T/T_c-1)c_1^2-\widetilde{J}_fc_1^4 a_{1}
\end{split}
\end{equation}
with $a_1=\frac14\int_{0}^{1}\rho_1^4u^{9-z} du|_{T=T_c}>0$ and,
\begin{equation}\label{expandN2}
N\simeq -\lambda^2c_1N_1/2.
\end{equation}
Putting \eqref{expandN2} into~\eqref{expand3}, we can obtain,

\begin{equation}\label{expand4}
\begin{split}
\widetilde{\Omega}(T,c_n)&\simeq\widetilde{\Omega}(T,N)\\
&\simeq \frac{2a_0}{\lambda^2N_1^2}(T/T_c-1)N^2+\frac{-16\widetilde{J}_fa_1}{\lambda^6N_1^4}N^4.
\end{split}
\end{equation}
It easy to see that this is just the Ginzburg-Landau (GL) theory of ferromagnetic model.
Based on the grand thermodynamic potential in Eq.~\eqref{expand4}, we can obtain the expression of magnetic moment in the ferromagnetic phase as,
\begin{equation}\label{onshellN}
N/\lambda^2=\sqrt{\frac{N_1^2a_0}{-16\widetilde{J}_fa_1}}(1-T/T_c)^{1/2}.
\end{equation}
This just confirms the critical behavior obtained in the numerical calculations and the critical exponent $1/2$ is an exact result. In that follows, we will compute all the coefficients  appearing in \eqref{onshellN} and compare them with the numerical ones.

\subsection{Spontaneous magnetization}
Let us first compute $N_1$ and $a_1$.  For this we have to first  find  the eigenfunction $\rho_1$, which is the solution of,
\begin{equation}\label{eqrho1}
-\frac{d}{du}\left[u^{3-z}f(u)\frac{d\rho_n}{du}\right]+q(u)\rho_n=0
\end{equation}
at $T=T_c$ with the conditions,
\begin{equation}\label{cond1}
\rho_1(1)=1,~~~\rho_{1+}=0.
\end{equation}
For convenience, here we do not assume that $\{\rho_n\}$ form an unit base. Thus we have,
\begin{equation}\label{N1s}
N_1=\frac1{C_1}\int_{0}^{1}\rho_1u^{1-z}du,~~a_1=\frac1{4C_1^4}\int_{0}^{1} u^{9-z}\rho_1^4 du,
\end{equation}
here $C_1$ is the normalization coefficient and
\begin{equation}\label{eqC}
C_1^2=\langle\rho_1,\rho_1\rangle=\int_{0}^{1} \omega\rho_1^2du.
\end{equation}
We have
\begin{equation}\label{onshellN2}
N^2/\mu_c^2=\frac{N_1^2a_0}{-16\widetilde{J}_fa_1\mu_c^2}(1-T/T_c)\simeq a_2(1-T/T_c).
\end{equation}
To clarify that the results are independent of the specific form of weight function, we choose $k=3, 4$ as two examples. From Tab.~\ref{tab:tc31} we see that different weight functions give different values for $N_{1}$, $a_{1}$ and $a_{0}$, but the same value for the magnetic moment $N$(up to a numerical error).

The value of $a_0$ can also be obtained directly by solving ODE~\eqref{SL1}. In the region near the critical temperature, we assume $\lambda_1=a_0(T/T_c-1)$. Note that all quantities in~\eqref{SL1} are the functions of temperature, thus taking derivative with respect to $T$ and evaluating at $T=T_c$, we get,
\begin{equation}\label{DSL1}
\frac{d\widehat{P}}{dT}\rho_1+\widehat{P}\frac{d\rho_1}{dT}=\frac{a_0}{T_c}\rho_1.
\end{equation}
Here $\rho_1$ is the eigenfunction of~\eqref{eqrho1}. Now treat $\rho_T=\frac{d\rho_1}{dT}$ as an unknown function to be solved, then the task to find $a_0$ becomes to solve a non-homogenous eigenvalue problem,
\begin{equation}\label{DSL2}
\widehat{P}\rho_T=\left[\frac{a_0}{T_c}-\frac{d\widehat{P}}{dT}\right]\rho_1.
\end{equation}
At the AdS boundary, $\rho_T$ has the same asymptotic behavior as~\eqref{asym1}, thus we can impose the boundary conditions as
\begin{equation}\label{bound}
|\rho_T(1)|<\infty,~\rho_{T+}=0.
\end{equation}
We find that $\rho_T\in D$. We then use the basis $\{\rho_n\}$ to expand $\rho_T$, i.e.,
\begin{equation}\label{expanda1}
\rho_T=\sum_{n=1}^\infty \frac{d_n}{C_n}\rho_n.
\end{equation}
Here $C_n$ are the modules of $\rho_n$. Using the fact $\lambda_1=0$ at $T=T_c$ and
\begin{equation}\label{expanda2}
\begin{split}
\langle C_1^{-1}\rho_1,\widehat{P}\rho_T\rangle&=\sum_{n=1}^\infty d_n\langle C_1^{-1}\rho_1,C_n^{-1}\widehat{P}\rho_n\rangle\\
&=\sum_{n=1}^\infty d_n\lambda_n\delta_{1n}=d_1\lambda_1=0.
\end{split}
\end{equation}
we have,
\begin{equation}\label{DSL3}
\begin{split}
\langle \rho_1,\left[\frac{a_0}{T_c}-\frac{d\widehat{P}}{dT}\right]\rho_1\rangle&=\frac{a_0C_1^2}{T_c}-\int_{0}^{1} \omega\rho_1\frac{d\widehat{P}}{dT}\rho_1 dz\\
&=\frac{a_0C_1^2}{T_c}-\int_{0}^{1}\rho_1\frac{d\widehat{L}}{dT}\rho_1 du=0.
\end{split}
\end{equation}
Furthermore we get,
\begin{equation}\label{DSL4a}
a_0=\frac{T_c}{C_1^2}\int_{0}^{1} dz\rho_1\frac{d\widehat{L}}{dT}\rho_1.
\end{equation}
This expression is valid in canonical ensemble with fixed charge density $\sigma=1$. It is very useful to find its equivalent form in the case with fixed $r_h=1$, since it is convenient when we perform numerical computation. If we fix $r_h=1$, the shooting parameter is chemical potential $\mu$. The relation between temperature in canonical ensemble and charge density is given by,
\begin{equation}\label{relTmu}
T=\frac{z+2}{4\pi\sigma^{z/2}}.
\end{equation}
Thus the expression \eqref{DSL4a} can be rewritten as
\begin{equation}\label{DSL4}
\begin{split}
a_0&=\frac{T_c}{C_1^2}\frac{d\mu}{dT}\int_{0}^{1} du\rho_1\frac{d\widehat{L}}{d\mu}\rho_1 \\
&=-\frac{4\mu_c^3\pi T_c}{(2+z)C_1^2}\int_{0}^{1} du\rho_1\left.\frac{d\widehat{L}}{d\mu}\rho_1\right|_{\mu=\mu_c} .
\end{split}
\end{equation}

\subsection{Susceptibility}

When $B\neq0$, the susceptibility for $T>T_c$ is defined as,
\begin{equation}\label{sus1}
\chi=\lim_{B\rightarrow0}\left(\frac{\partial N}{\partial B}\right)_T.
\end{equation}
In the case with $T>T_c$ and $B\rightarrow0$, we can neglect the non-linear term, i.e., setting $\widetilde{J}_f=0$.  The solution of equation~\eqref{rhop} can be expressed as,
\begin{equation}
\rho=\sum_{n=1}^\infty c_n\rho_n+\frac{B}{m^2},
\end{equation}
Taking into account the equation~\eqref{rhop} with $\widetilde{J}_f=0$, we have
\begin{equation}\label{rhop2}
\begin{split}
0&=\widehat{L}\rho-Bu^{1-z}=\sum_{l=1}^\infty c_l\widehat{L}\rho_l+\frac{4 BJ \mu^2u^{5-z}(2-z)^2}{m^6}\\
&=\sum_{l=1}^\infty c_lC_l^{-1}\lambda_l\omega\rho_l-\frac{4 BJ \mu^2z^4}{m^6}.
\end{split}
\end{equation}
Multiplying a factor $\rho_n/C_n$ and integrating the above equation from 0 to 1, we can obtain
\begin{equation}\label{expandB1}
c_n=-\frac{B\gamma_n}{\lambda_n},~~\text{with}~\gamma_n=\int_{0}^{1}\frac{4J \mu^2u^{5-z}(2-z)^2}{C_n m^6}\rho_ndu.
\end{equation}
Thus we can get the magnetic moment density as
\begin{equation}\label{meg1}
N/\lambda^2=-\int_{0}^{1}\frac{Bu^{1-z}}{2m^2}du+B\sum_{n=1}^\infty \frac{\gamma_nN_n}{2\lambda_n},
\end{equation}
and the magnetic susceptibility
\begin{equation}\label{chi1}
\chi/\lambda^2=-\int_{0}^{1}\frac{u^{1-z}}{2m^2}du-\sum_{n=1}^\infty \frac{\gamma_nN_n}{2\lambda_n}.
\end{equation}
When $T\rightarrow T_c^+$, we have $\lambda_1=a_0(T/T_c-1)\rightarrow0^+$. Thus $\chi$ is dominated by the first term in the summation of~\eqref{chi1} and its inverse can be expressed as

\begin{equation}\label{chi2}
\lambda^2\chi^{-1}/\mu_c^{\frac{2-z}{2}}=\frac{2a_0}{\mu_c^{\frac{2-z}{2}}}{\gamma_1N_1}(T/T_c-1),~~\text{as}~T\rightarrow T_c^+.
\end{equation}
In the case of $m^2=-J=1/8$, for example, we have $\lambda^2\chi^{-1}/\mu_c\simeq2.996(T/T_c-1)$, which is very close to our numerical result $\lambda^2\chi^{-1}/\mu_c\simeq2.966(T/T_c-1)$ given in the numerical calculation when we take $z=1$ in the case of 4D spacetime. But from Tab.~\ref{tab:tc31}, it easy to see that the magnetic moment and susceptibility have a great gap when we compare the analytical method with numerical calculation, especially for the case of 5D Lifshitz spacetime.

\begin{figure}
\begin{center}
\includegraphics[width=0.4\textwidth]{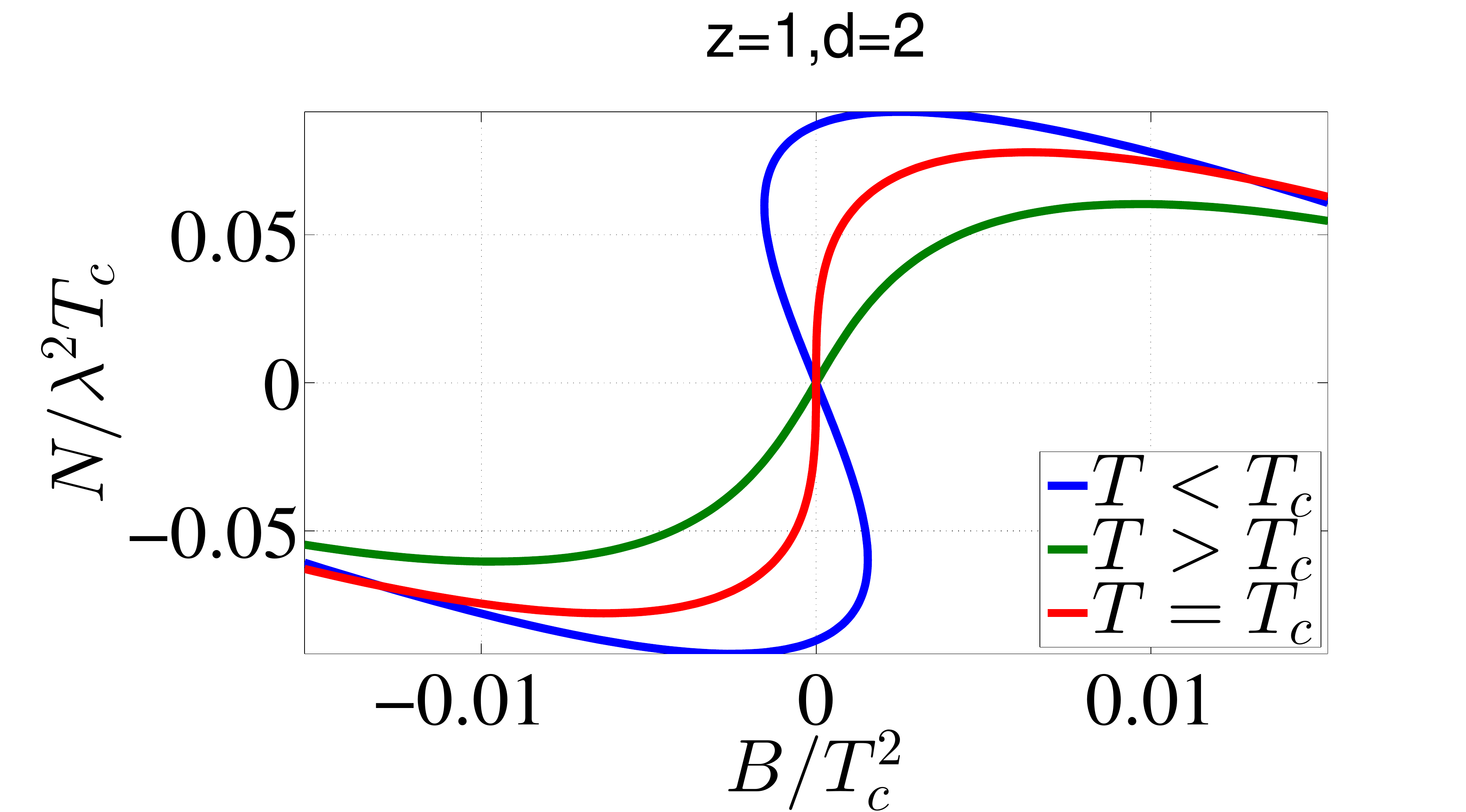}
\includegraphics[width=0.4\textwidth]{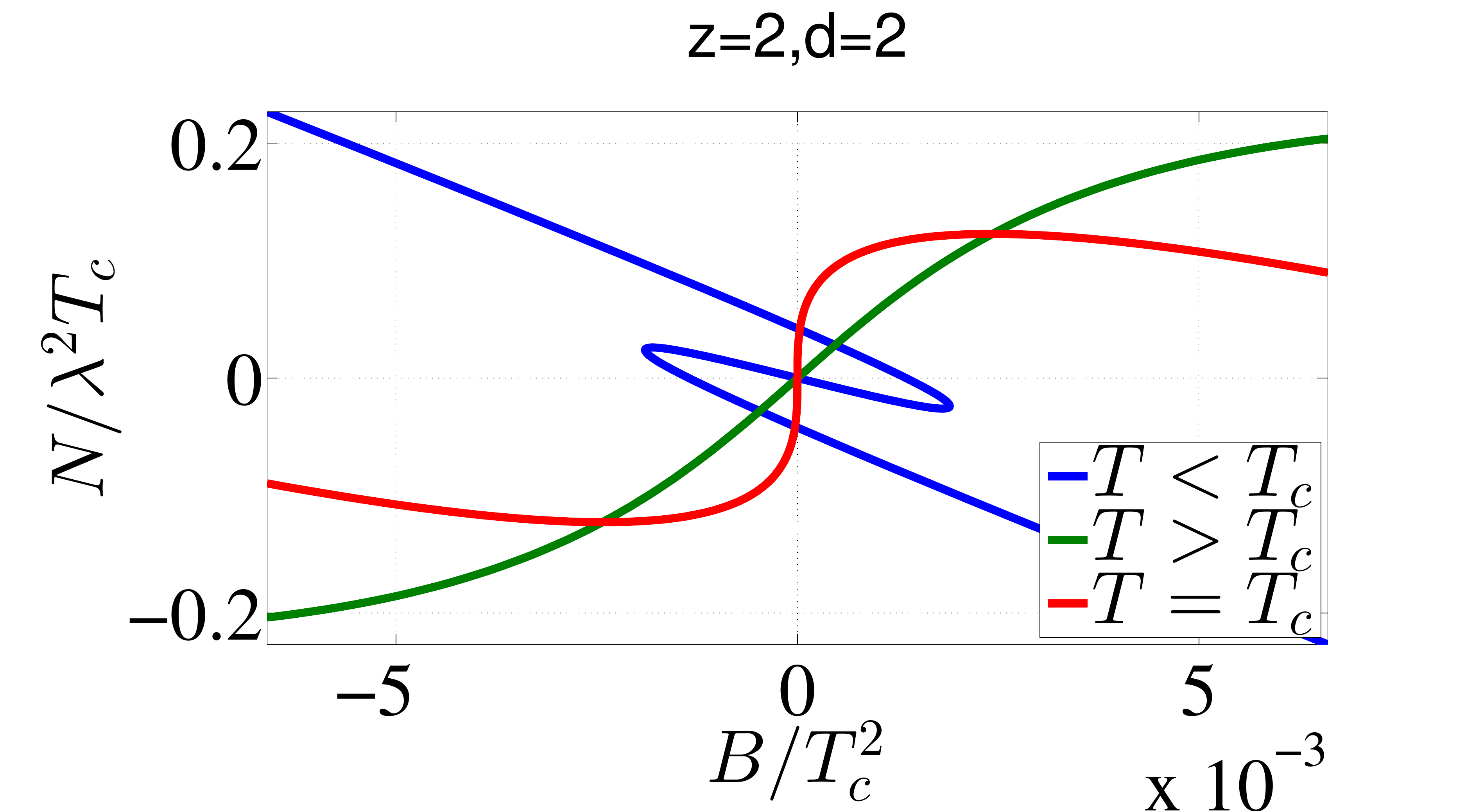}
\includegraphics[width=0.4\textwidth]{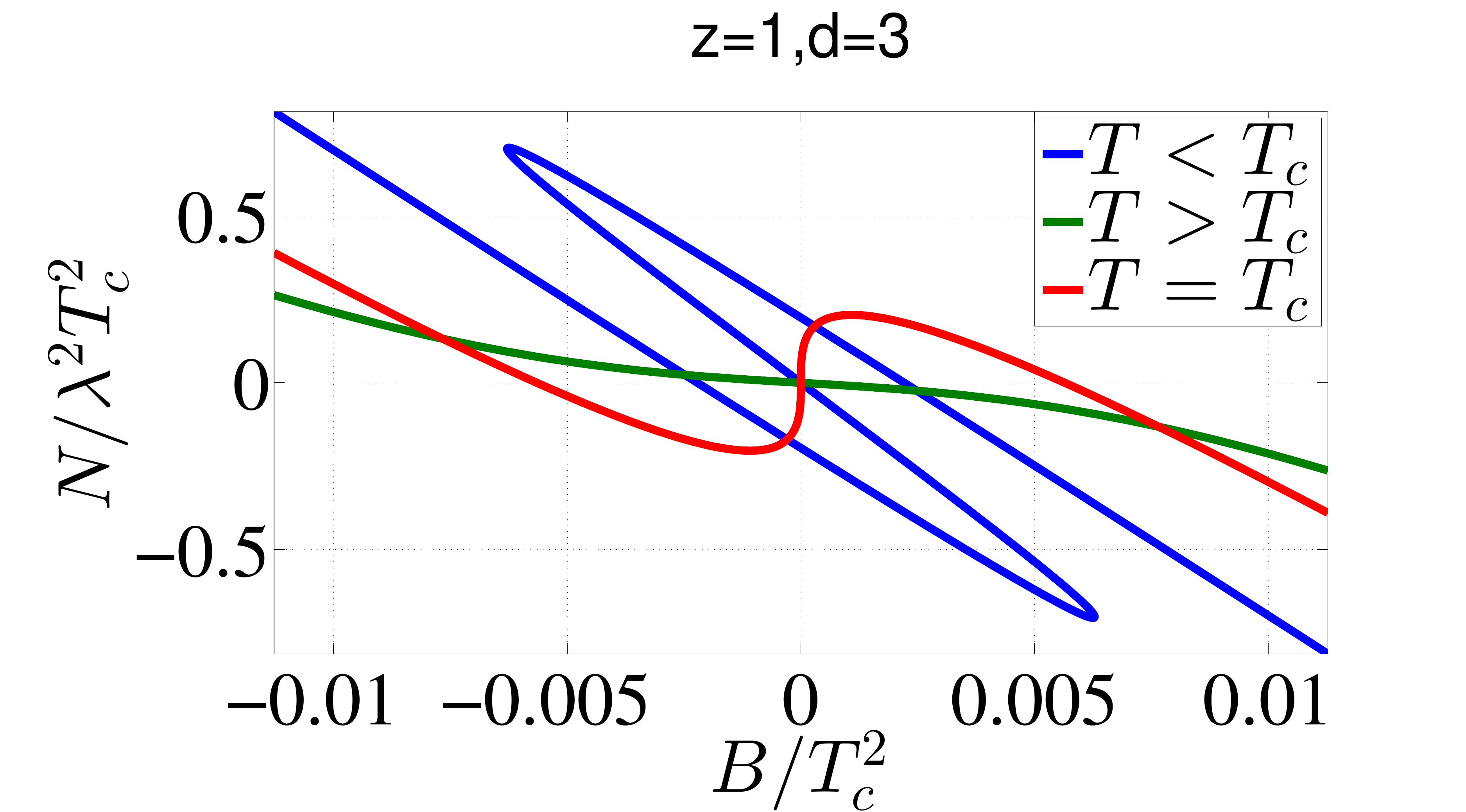}
\includegraphics[width=0.4\textwidth]{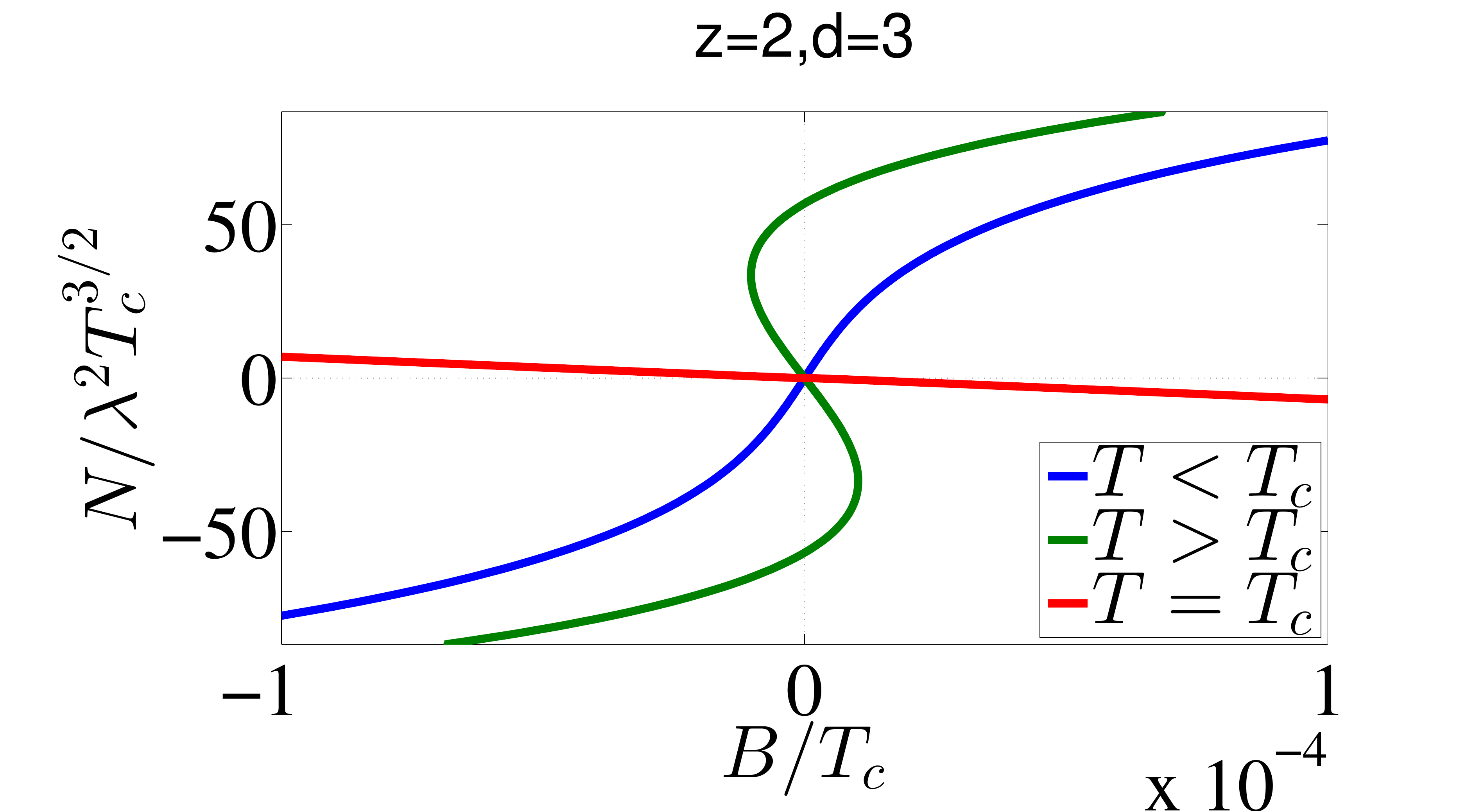}
\includegraphics[width=0.4\textwidth]{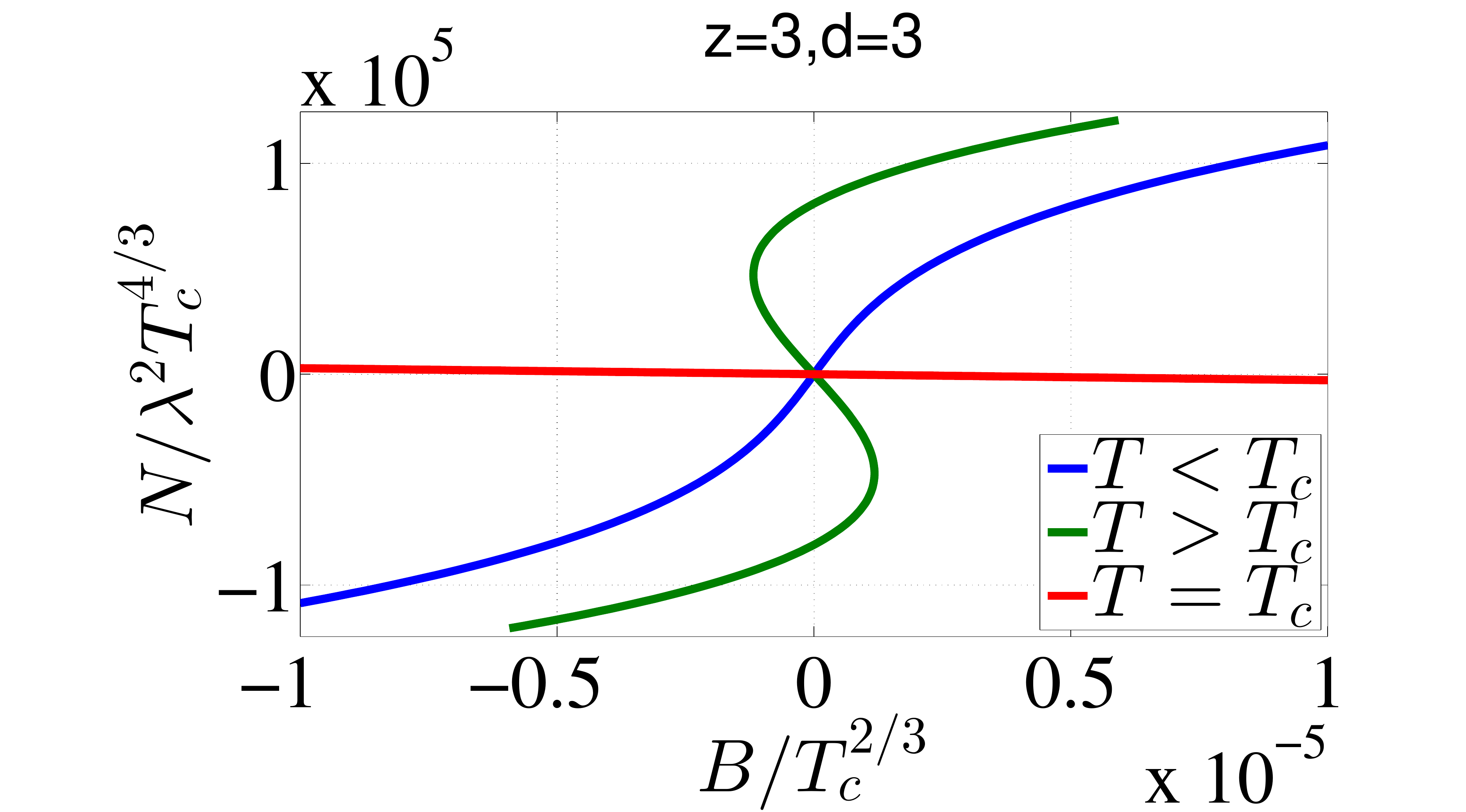}
\caption{The relation between magnetic moment density N
and external magnetic field B in the cases of $T = 1.05Tc$, $T =
0.9T_c$ and $T=T_c$, respectively}
\label{relBN}
\end{center}
\end{figure}

Below let us move to the case with $B\neq0$. In this case from \eqref{rhop} we have,
\begin{equation}\label{eqB1}
\int_{0}^{1}\rho_n(\omega\widehat{P}\rho-B u^{1-z}-\widetilde{J}_f\rho^3u^{9-z})du=0.
\end{equation}
According to \eqref{exprho1}, we can rewrite it as
\begin{equation}\label{eqBN}
\int_{0}^{1}\rho_n(\omega\widehat{P}\widetilde{\rho}-B[u^{1-z}-\frac{q(u)}{m^2}]-\widetilde{J}_f\rho^3u^{9-z})du=0.
\end{equation}
Using the expansion expression~\eqref{exprho1}, we have,
\begin{equation}\label{eqBN2}
c_nC_n^2\lambda_n+B\gamma_n-\int_{0}^{1}\rho_n\widetilde{J}_f\rho^3u^{9-z}du=0,~~n=1,2,\cdots.
\end{equation}
For convenience, we assume that $\{\rho_n\}$ is an unit base, i.e., $C_n$=1. Equation~\eqref{eqBN2} is equivalent to \eqref{rhop} if we take all the terms in \eqref{exprho1} into account. In the case of $T\rightarrow T_c^-$, assuming that  the first term in \eqref{exprho1} dominates only, i.e., $|c_1|\gg c_n$ for $n\geq2$ in \eqref{coeffN}, we get
\begin{equation}
\label{eqN2}
N/\lambda^2=-\frac{B}{2m^2}\int_{0}^{1}u^{1-z}du -c_1N_1/2.
\end{equation}
Taking $n=1$ in \eqref{eqBN2}, we have,
\begin{equation}\label{eqBN3}
c_1\lambda_1+B\gamma_1-c_1^3\widetilde{J}_f\int_{0}^{1}\rho_1^4u^{9-z}du=c_1\lambda_1+B\gamma_1-4c_1^3\widetilde{J}_fa_1=0.
\end{equation}
For a given temperature $T\rightarrow T_c$, we can combine~\eqref{eqN2} with~\eqref{eqBN3} to obtain a relation between  external magnetic field $B$ and magnetic moment $N$. Figure~\ref{relBN} shows the results with $T=1.05T_c, T=0.9T_c$ and $T=T_c$, respectively, in the case of $m^2=-J=1/8$. We see that it is very similar to what we have obtained in~\cite{Cai:2014oca}, particularly for the case of $z=1,d=2$.
In the case of $T<T_{c}$, when the external field continuously changes between $-B_{max}$ and $B_{max}$ periodically, the metastable states of magnetic moment can appear. Thus we see a hysteresis loop in the single magnetic domain. Furthermore from Fig.~\ref{relBN}, it is easy to see that the magnetic moment is not single valued with a general $z$ and D, and the Lifshitz dynamical exponent $z$ has an effect on the hysteresis loop quantitatively. Particularly, for the cases of $z=2$ in 4D and $z=3$ in 5D spacetime, the period of the external field $B$ is shorter and the value of the magnetic moment $N$ will become bigger than ones for the cases of other $z$ in 4D and 5D spacetime.

\section{Summary and discussion}
In summary, we have numerically and analytically investigated the holographic paramagnetism/ferromagnetism phase transition model in the 4D and 5D Lifshitz black holes in the probe limit by introducing a massive 2-form field coupled to the background Maxwell field, and obtained the effects of the dynamical exponent z on the holographic paramagnetism/ferromagnetism phase transition. Our results are concluded as follows.

We have obtained the critical temperature $T_{c}$ firstly if the
model parameters are in some suitable region and then typically
plotted the magnetic moment and the inverse susceptibility density
as a function of the temperature. The results show that, in the
case without external magnetic field, the improving of dynamical
exponent $z$ results in the increase of $T_{c}$, which implies
that the increasing anisotropy between space and time enhances the
phase transition. Especially, for the case of 5D Lifshitz
spacetime, the value of magnetic moment changes obviously compared
with the 4D spacetime. In the vicinity of the critical point,
however, the behavior of the magnetic moment is always as
$\sim(1-T/T_{c})^{1/2}$, regardless of the values of $z$ and D,
which is in agreement with the result from mean field theory. And
the DC resistivity is not relevant to the dynamical exponent $z$
qualitatively, though in this probe limit it is suppressed by
spontaneous magnetization and shows a metallic behavior. But the
value of DC resistivity is influenced when the sample gets
cooling, i.e., the bigger the value of $z$, the bigger the DC
resistivity although it decreases with the temperature. Moreover,
in the presence of the external magnetic field, the inverse
magnetic susceptibility near the critical point behaves as
$\sim(T/T_{c}-1)$ in all cases, which satisfies the Cure-Weiss
law.

Furthermore, by semi-analytic method we have calculated the
magnetic moment and static magnetic susceptibility, and obtained
the relation between external magnetic field $B$ and magnetic
moment $N$ near the critical temperature. And we have observed the
hysteresis loop in the single magnetic domain when the external
field continuously changes between the maximum and minimum values
periodically with a general $z$ or D. But for the fixed value of
D, the increase of the dynamical exponent $z$ could result in
shortening the period of the external magnetic field. In addition,
the transformation period is smaller in the 5D case than one in
the 4D case.

Note that in this paper we only worked on the probe limit by neglecting the backreaction of the matter fields. Although the probe limit can reveal some significant properties of holographic ferromagnetic phase transition, maybe the order of the phase transition could be changed once the backreaction is taken into consideration, and some new phases could emerge. Therefore, it is interesting to study the influence of the backreaction of matter field to the Lifshitz background and to see whether there are some new features beyond the probe limit, which will be our research work in the near future.

\section*{Acknowledgments}
We would like to thank Prof. R. G. Cai and Dr. R. Q. Yang for their helpful discussions and comments. This work is supported by the National Natural Science Foundation of China (Grant Nos. 11175077, 11575075) and the Joint Specialized Research Fund for the Doctoral Program of Higher Education, Ministry of Education, China (Grant No. 20122136110002).

\end{document}